\documentclass[final,3p,times,twocolumn]{elsarticle}
\usepackage{amsmath}
\usepackage{amsthm}
\usepackage{braket}
\usepackage{subfigure} 
\usepackage{arydshln}
\usepackage{mathrsfs}
\usepackage{multicol}
\usepackage{widetext}
\usepackage{mathtools, cuted}
\usepackage{amssymb}
\usepackage{multirow,booktabs}

\biboptions{numbers,sort&compress}

\newlength{\halfpagewidth}
\divide\halfpagewidth by 1

\newcounter{bla}

\journal{Computer Physics Communications}

\begin{document}

\begin{frontmatter}



\title{RTGW2020: A powerful implementation of DFT + Gutzwiller method}


\author[a,b]{Shiyu Peng}
\author[a,b,c]{Hongming Weng}
\author[d]{Xi Dai\corref{author}}

\cortext[author] {Corresponding author.\\\textit{E-mail:} daix@ust.hk}
\address[a]{Beijing National Laboratory for Condensed Matter Physics, and Institute of Physics, Chinese Academy of Sciences, Beijing 100190, China}
\address[b]{University of Chinese Academy of Sciences, Beijing 100049, China}
\address[c]{Songshan Lake Materials Laboratory, Dongguan, Guangdong 523808, China}
\address[d]{Department of Physics, Hong Kong University of Science and Technology, Clear Water Bay, Kowloon 999077, Hong Kong}

\begin{abstract}
In the present paper, we propose an efficient numerical scheme for Gutzwiller method for multi-band Hubbard models with general onsite Coulomb interaction. Following the basic idea of \textit{Deng et al.} [Phys. Rev. B 79, 075114 (2009)] and extensions by \textit{Lanata et al.} [Phys. Rev. B 85, 035133 (2012)], the ground state is variationally determined through optimizing the total energy with respect to the variational single particle density matrix ($\boldsymbol{n^0}$), which is called "outer loop". In the corresponding "inner loop" where $\boldsymbol{n^0}$ is fixed, the non-interacting wave function and the parameters contained in the Gutzwiller projector are determined by a two-step iterative approach. All derivatives of the implementation process have been analytically derived, which allows us to apply some advanced minimization or root-searching algorithms for both the inner and outer loops leading to the highly efficient convergence. In addition, an atomic diagonalization method taking the point group symmetry into account has been developed for the customized design of the Gutzwiller projector, making it convenient to explore many interesting orders at a lower cost of computation. As benchmarks, several different types of correlated models have been studied utilizing the proposed method, which are in perfect agreement with the previous results by DMFT and multi-orbital slave-boson mean field method. Compared with the linear mixing method, the newton method with analytical derivatives shows much faster convergence for the inner loop. As for the outer loop, the minimization using analytical derivatives also shows much better stability and efficiency compared with that using numerical derivatives.
\end{abstract}


\end{frontmatter}

\begin{small}

\section{Introduction}
\label{itdct}
How to perform efficient and accurate calculations for the strongly correlated materials, where various exotic phenomena like superconductivity, magnetism as well as heavy fermions behavior emerges, remains a great challenge nowadays. The density-functional theory (DFT) \cite{hohenberg1964inhomogeneous,kohn1965self} with local density approximation (LDA) \cite{gunnarsson1976exchange} or the generalized gradient approximation (GGA) \cite{perdew1996generalized} has gained great success in many real matter systems such as simple metals and semiconductors. However, due to the lack of proper consideration of the many-body correlation effects among the electrons, it fails when applied to the strongly correlated materials, for example, SrVO$_3$, CoO, SmB$_6$ and URu$_2$Si$_2$. Over the last three decades, several methods are proposed to be supplemented to LDA and GGA, such as LDA $+$ U \cite{anisimov1991band,anisimov1993density,solovyev1994corrected,liechtenstein1995density}, LDA $+$ dynamical mean-field theory (DMFT) \cite{kotliar2006electronic} as well as the GW method \cite{aryasetiawan1998gw}. However, the methods mentioned above are either inaccurate or time-consuming for most $3d$ and $4-5f$ strongly correlated systems. In this paper, we introduce a new computational scheme of Gutzwiller variational method, which can be easily combined with the LDA method to study the correlated materials and is proven to be well balanced between accuracy and efficiency for the ground state calculations.

The Gutzwiller variational method is named after Martin Gutzwiller who proposed the Gutzwiller trial wave function (GWF) as a variational ansatz \cite{gutzwiller1963effect,gutzwiller1964effect} to study the itinerant ferromagnetism in 3d transition metals characterized by the Hubbard model. The GWF is defined by a so-called Gutzwiller projector acting on the non-interacting ground state, through which the weights of different atomic configurations are tuned to minimize the total energy. With the help of a further Gutzwiller approximation (GA) \cite{gutzwiller1965correlation}, which is later proven to be exact in the infinite dimensions, the physical observable can be evaluated using the Gutzwiller wave function. Over the last decade, there have been a series of progress \cite{vollhardt1984normal,vollhardt1990gutzwiller,metzner1987ground,metzner1988analytic,gebhard1987correlation,gebhard1988correlation,metzner1989correlated,metzner1989variational,bunemann1997gutzwiller,bunemann1997generalized,bunemann2000multi,attaccalite2003properties} on this method, which brings it from an analytical algorithm which is only suitable for one- or two-band Hubbard model to a numerical method for the study of realistic materials. In particular, Deng \textit{et al.} \cite{deng2009local} and Ho \textit{et al.} \cite{ho2008gutzwiller} proposed almost simultaneously to combine the DFT and the Gutzwiller method (LDA $+$ G) to make it a new computational method for practical calculations of strongly correlated materials. The LDA $+$ G method consists of two major steps. First, the regular LDA calculation will be carried out to obtain the "non-interacting" electronic Hamiltonian including both correlated and non-correlated energy bands. Second, local Hubbard type interaction will be implemented among the so called "correlated local orbitals" such as 3d and 4f orbitals, where the strong correlation effect takes place, and treated by the Gutzwiller variational method. In the early version of the LDA $+$ G method develped in our group, only the density-density type of Coulomb interaction can be treated, which limits its application. Later, \textit{Lanata} \textit{et al.} \cite{lanata2012efficient} extended it to general interaction including rotational invariant Hund's couplings which are important for the cases of multiple active local orbitals. Another shortcoming of the current numerical scheme for the Guztwiller method is that it is based on the iterative process suffering from the poor converging property and low computational efficiency, which needs to be improved.

In this paper, we firstly review the main procedure of the Gutzwiller method as well as the notation of $\boldsymbol{\phi}$ matrix proposed by \textit{Lanata} \textit{et al.} \cite{lanata2008fermi}. Then we reformulate the Gutzwiller total energy to be a functional of three classes of variational parameters, the reduced single particle density matrix within the local orbitals, the non-interacting ground state and the parameters contained in the Gutzwiller projector describing the adjustment of the local atomic configurations. The entire minimization process will be divided into two different levels. The top most level is the energy minimization with respect to the variational single particle density matrix, which is called "outer loop" of the entire Gutzwiller minimization. The second level of minimization is called "inner loop", which corresponds to minimize total energy for given single particle density matrix respect to all other variational parameters. In order to improve the efficiency of root-searching and optimization used in the above mentioned steps , we derive all the analytical derivatives of the real implementation process by means of the perturbation theory and the chain rule. In this new implementation of the Gutzwiller variational method, we apply the full point group symmetry analysis to reduce the number of the independent variational parameters in the Gutzwiller projectors. As will be shown below, the symmetry consideration can not only raise the computational efficiency for the paramagnetic state but also provide a systematic way for the studies of the spontaneous symmetry breaking phase. In order to show the accuracy and efficiency of this new implementation, two simple models are studied for the benchmarks with the old implementation of the Gutzwiller method, multi-orbital slave-boson mean field method as well as DMFT. The comparison between analytical derivatives and numerical derivatives by finite difference is also shown.

The rest of this paper is organized as follows.  Sec. \ref{mtd} includes most of the theoretical content. After the review of some basic formulas and the notation of $\boldsymbol{\phi}$ matrix in the first two parts, the energy functional of LDA $+$ G is then discussed. In the fourth part of Sec. \ref{mtd},  we demonstrate the variational scheme of the constrained functional. After that, how to optimize the energy functional through derivatives within the new scheme is shown. In the last part of Sec. \ref{mtd}, the interface to \textit{ab initio} codes is introduced. In Sec. \ref{gvp}, we describe how to take the point group symmetry into consideration when constructing the Gutzwiller projector. Next, in Sec. \ref{imple}, how to implement the DFT + G method is hierarchically described. In Sec. \ref{bcmk}, we show the efficiency and accuracy of this new implementation for two lattice models. Conclusions are summarized in Sec. \ref{ccls} and some detailed derivations are given in Appendix.

\section{Method}
\label{mtd}
\subsection*{Symbols}
To avoid ambiguousness, we first introduce the notation adopted in this paper.
\begin{itemize}
\item $i, j$  : site indices. \\
\item $\alpha, \beta$  : indices for spin and orbital. \\
\item $\hat{O}, \hat{H}, \cdots$  : quantum operators. \\
\item $\boldsymbol{F}, \boldsymbol{a}, \boldsymbol{n^0}, \boldsymbol{\phi}, \cdots$  : generic tensors including vectors and matrices. \\
\item $F_{mn}, a_m, n^0_\alpha, \phi_{mn}, \cdots$  : components of  general tensors.\\
\item $\mathbb{R}, \mathbb{N}, \mathbb{F}, \cdots$  : matrices under the notation of $\boldsymbol{\phi}$ matrix.\\
\item $\mathscr{E}$ : energy functional.\\
\item $\ket{I}, \ket{J}$ : the  Fock bases.\\
\item $\ket{\Gamma}$ : the atomic eigenstates.
\end{itemize}

\subsection{The Gutzwiller method}

Let us start from the general Hubbard model which reads
\begin{equation}
\hat{H} = \sum_{i\ne j;\alpha,\beta} t^{\alpha\beta}_{ij} \hat{c}^\dag_{i\alpha} \hat{c}_{j\beta} +  \sum_i \hat{H}^{at}_i
\end{equation}
where i (j) is the site index of lattice and $\alpha$ ($\beta$) labels both spin and orbital. $\hat{H}_i$ denotes all the total onsite term including intra-site general Coulomb interaction and spin-orbital coupling as well as the crystal field.

Professor Martin Gutzwiller first introduced a famous Gutzwiller trial wave function (GWF) which reads

\begin{equation}
\ket{G} =\hat{P}\ket{0} \equiv \prod_i \hat{P}_i \ket{0} \quad .
\end{equation}
where $\ket{0}$ is the wave function for the free fermions. The ground state of the Hubbard model is variationally determined by optimizing the local projector $\hat{P}_i$ whose general form can be defined as 
\begin{equation}\label{pjt}
\hat{P}_i = \sum_{\Gamma,\Gamma'} \lambda_{i;\Gamma\Gamma'}\ket{i,\Gamma} \bra{i,\Gamma'} \quad .
\end{equation}
where $\ket{i,\Gamma}$ represents the atomic eigenstate on site $i$. The local projector $\hat{P}_i$ plays a central role in the Gutzwiller method which is used to adjust the weights ($\lambda_{i;\Gamma\Gamma'}$) of different configurations to achieve the total energy minimum. The concrete form of $\hat{P}_i$ can be designed with respect to the local crystal symmetry as will explain in more details later.

The total energy is obtained by evaluating the Hamiltonian using the GWF
\begin{equation}
\mathscr{E}[\ket{0},\boldsymbol{\lambda}] = \bra{G}\hat{H}\ket{G} = \bra{0}\hat{P}^\dag \hat{H} \hat{P}\ket{0}
\end{equation}
where $\boldsymbol{\lambda}$ is a matrix describing the coefficients ($\lambda_{i;\Gamma\Gamma'}$) in the Gutzwiller projector defined in Eq. (\ref{pjt}) and the site index $i$ is dropped off here for translational invariant systems as well.

The total energy should be minimized under two Gutzwiller constraints (GC) which are
\begin{equation}\label{cst}
\begin{split}
\bra{0}\hat{P}^\dag_i\hat{P}_i\ket{0}&=1 \\
\bra{0}\hat{P}^\dag_i \hat{P}_i \hat{n}_{i\alpha} \ket{0} &= \bra{0} \hat{n}_{i\alpha} \ket{0} \quad .
\end{split}
\end{equation}
The first one is nothing but the normalization of GWF and the second one is a natural result of Gutzwiller approximation (GA) \cite{fabrizio2007gutzwiller}.  These two constraints make it possible to get analytically exact expectation value of observables in infinite dimension systems.

A similar quantity is the expectation value of observables,
\begin{equation}\label{ob1}
\bra{G}\hat{O}_i\ket{G} = \bra{0}\hat{P}^\dag \hat{O}_i \hat{P}\ket{0}
\end{equation}
If the system has infinite coordinate and together with both GC, Eq. (\ref{ob1}) can be reduced to 
\begin{equation}\label{ob2}
\bra{G}\hat{O}_i\ket{G} = \bra{0}\hat{P}_i^\dag \hat{O}_i \hat{P}_i\ket{0} \quad .
\end{equation}
Although strictly speaking the above equation is only valid for the infinite dimension systems \cite{fabrizio2007gutzwiller}, it is a good approximation to study the finite dimension systems as well. 

Michele Fabrizio \cite{fabrizio2007gutzwiller} has clarified the difference and relation between Eq (\ref{ob2}) and the second constraint of Eq (\ref{cst}). Generally they are different unless $\hat{P}_i$ commutes with $\hat{O}_i$ in very limited cases. For example, the projector is diagonal and the onsite interaction is taken as a density-density form.

\subsection{$\boldsymbol{\phi}$ matrix}
The implementation of Gutzwiller method depends heavily on the operation of large matrices or tensors. Therefore, it is convenient for us to formulate most of the quantities (local observables, energy functional and constraints) in $\boldsymbol{\phi}$ matrix introduced originally in Ref. \cite{lanata2008fermi}. Although many formulas in this subsection have been derived in a similar manner with \textit{Lanata et al.} under the so-called mixed-basis in Ref. \cite{lanata2012efficient}, we will derive these formulas in natural basis again to make the present paper self-contained.

First let us introduce the natural basis which is adopted in this paper to make single particle density matrix diagonal,
\begin{equation}
\bra{0}\hat{c}^\dag_{i\alpha} \hat{c}_{i\beta}\ket{0} = n^0_{i,\alpha} \delta_{\alpha\beta}
\end{equation}
where $\boldsymbol{n^0}$ is the local particle number here which will be redefined as so-called variational single particle density matrix afterwards. Although the choice of basis is not unique, however, the natural basis has been proven to be very useful for practical implementation of the Gutzwiller method, especially for evaluating operators in many-body space. Take the many-body projector "$\ket{i,I}\bra{i,I'}$" for example,
\begin{equation}\label{m0fock}
\braket{0|i,I}\braket{i,I'|0} = \delta_{II'}\prod_{\alpha} (n^0_{i,\alpha})^{n^I_{i,\alpha}}(1-n^0_{i,\alpha})^{1-n^I_{i,\alpha}} \equiv m^0_{i,I} \delta_{II'}
\end{equation}
with $n^I_{i,\alpha} = \bra{i,I}\hat{n}_{i,\alpha}\ket{i,I}$ equals 1 or 0 depending on whether the $\alpha$-th orbital of $\ket{i,I}$ is occupied or not. Under more generic basis, the evaluation should be the summation of contractions in all possible ways.

The $\boldsymbol{\phi}$ matrix is a reformulation of the local Gutzwiller projector on site $i$ and can be further expanded by a set of independent local matrices $\{\boldsymbol{\phi^l_i}\}$ which make the description simpler,  
\begin{equation}\label{avec}
\boldsymbol{\phi_i} \equiv \boldsymbol{\lambda_i} \sqrt{\boldsymbol{m^0_i}} = \sum_{l=1}^{n_v} a_{i,l} \boldsymbol{\phi^l_i} \quad .
\end{equation}
where $\boldsymbol{m^0_i}$ is a diagonal matrix defined on the Fock space at the site $i$ whose elements are $m^0_{i,I}$ defined in Eq. (\ref{m0fock}). $n_v$ is the number of independent $\boldsymbol{\phi^l_i}$ matrices at the site $i$.

Any local observables can thus be represented in $\boldsymbol{\phi}_i$ matrix,
\begin{equation}\label{ob3}
\bra{G}\hat{O}_i\ket{G} = Tr(\boldsymbol{\phi}^\dag_i \boldsymbol{O}_i \boldsymbol{\phi}_i) = \boldsymbol{a}_i \mathbb{O}_i \boldsymbol{a}_i^T
\end{equation}
where 
\begin{equation}
\begin{split}
\mathbb{O}_{i,mn} &= Tr(\boldsymbol{\phi}^{m \dag}_i \boldsymbol{O}_i \boldsymbol{\phi}^n_i) \\
O_{i,\Gamma\Gamma'} &= \bra{i,\Gamma}\hat{O}_i\ket{i,\Gamma'} \\
\end{split}
\end{equation}
and $\boldsymbol{a}_i$ is a 1D array defined as $(a_{i,1},a_{i,2},\cdots,a_{i,n_v})$ while $\boldsymbol{a}^T_i$ is its transpose. 

The total energy consisting of kinetic energy ($E^{kin}$) and interaction energy ($E^a$) is derived in the literatures \cite{deng2009local, lanata2012efficient} as 
\begin{equation}\label{op1}
\begin{split}
\mathscr{E}[\ket{0},\boldsymbol{a}] & = E^{kin}[\ket{0},\boldsymbol{a}]  + E^a[\ket{0},\boldsymbol{a}] \\
& = \sum_{i\ne j;\alpha,\beta,\delta,\gamma} t^{\alpha\beta}_{ij}\mathcal{R}^\dag_{i,\alpha\gamma} \bra{0}\hat{c}^\dag_{i\gamma} \hat{c}_{j\delta}\ket{0} \mathcal{R}_{j,\delta\beta} +  \sum_i \boldsymbol{a}_i \mathbb{L}_i \boldsymbol{a}^T_i\\
\end{split}
\end{equation}
with the orbital renormalization factor,
\begin{equation}\label{r}
\mathcal{R}_{i,\delta\beta} = \frac{Tr(\boldsymbol{\phi}^\dag_i \boldsymbol{S}_{i\beta}  \boldsymbol{\phi}_i \boldsymbol{S}^\dag_{i\delta})}{\sqrt{n^0_{i,\delta}(1-n^0_{i,\delta})}} = \boldsymbol{a}_i \mathbb{R}^{\delta\beta}_i \boldsymbol{a}^T_i
\end{equation}
where $\boldsymbol{S}^\dag_{i\delta,II'} = \bra{i,I}c^\dag_{i,\delta}\ket{i,I'}$ and $\mathbb{R}^{\delta\beta}_{i,mn}=\frac{Tr(\boldsymbol{\phi}^{m \dag}_i \boldsymbol{S}_{i\beta}  \boldsymbol{\phi}^n_i \boldsymbol{S}^\dag_{i\delta})}{\sqrt{n^0_{i,\delta}(1-n^0_{i,\delta})}}$ which shows $\mathcal{R}$ is function of both $\boldsymbol{n^0}$ and $\boldsymbol{a}$ (or $\boldsymbol{\lambda}$). $\mathbb{L}_i$ is the local atomic Hamiltonian under the notation of $\boldsymbol{\phi}$ matrix. The $\boldsymbol{\mathcal{R}}$ matrix which is absent in Hartree Fock mean field approximation is a crucial result of the Gutzwiller method. The strong correlation effect within the localized orbitals take effect on kinetic terms through this factor. The derivation of $\boldsymbol{\mathcal{R}}$ can be found in the appendix.

Finally we also have the constraints reformulated as 
\begin{equation}
\begin{split}
Tr(\boldsymbol{\phi}^\dag_i \boldsymbol{\phi}_i)&=\boldsymbol{a}_i \mathbb{F}_i \boldsymbol{a}^T_i=1 \\
Tr(\boldsymbol{\phi}^\dag_i \boldsymbol{\phi}_i \boldsymbol{N}_{i\alpha}) &=\boldsymbol{a}_i \mathbb{N}_{i\alpha} \boldsymbol{a}^T_i =\bra{0} \hat{n}_{i,\alpha} \ket{0} \quad .
\end{split}
\end{equation}
with
\begin{equation}
\begin{split}
\mathbb{F}_{i,mn} &= Tr(\boldsymbol{\phi}^{m\dag}_i \boldsymbol{\phi}_i^n) \\
\mathbb{N}_{i\alpha,mn} &= Tr(\boldsymbol{\phi}^{m\dag}_i\boldsymbol{\phi}^n_i\boldsymbol{N}_{i\alpha})
\end{split}
\end{equation}
where $N_{i\alpha,\Gamma\Gamma'} = \bra{\Gamma} \hat{n}_{i,\alpha} \ket{\Gamma'}$.

As mentioned in Lanata's paper \cite{lanata2012efficient}, the matrices in blackboard bold font are unchanged during the inner iterations. Therefore, we can calculate them first and store them in the hard disk at the very beginning of the inner loop. For large systems or systems need many inner steps to achieve convergence, this strategy can save lots of time.

\subsection{Energy functional}
The key task of a variational problem is the total energy functional. How to optimize the functional depends on the way we construct it. Different perspectives of the functional construction lead to different independent variables taken as the variational parameters.

The total energy is treated originally as functional of both $\ket{0}$ and $\boldsymbol{a}$ shown in Eq. (\ref{op1}) in the previous section. However, the traditional iterative process based on this perspective introduced in \textit{Deng}'s paper is hard to converge in many cases. In this paper, the variables are expanded explicitly from $\{\ket{0},\boldsymbol{a}\}$ to $\{\boldsymbol{n^0},\ket{0},\boldsymbol{a}\}$ with several additional constraints on the new variable $\boldsymbol{n^0}$ termed as "variational single particle density matrix". In this way, the entire energy minimization can be implemented in two levels. The top most level dubbed as "outer loop" is the energy minimization respect to $\boldsymbol{n^0}$. Then in the second level, which is called "inner loop", the energy minimization is done respect to $\{\ket{0},\boldsymbol{a}\}$ with $\boldsymbol{n^0}$ fixed.

In the inner loop, the calculated single particle density matrix should equal to the given $\boldsymbol{n^0}$, which introduced a new constraint

\begin{equation}\label{cst1_n0}
n^F_{i,\alpha} \equiv \bra{0} \hat{n}_{i,\alpha} \ket{0} = n^0_{i,\alpha}, \quad \forall \alpha \in \{1,\cdots,n_{orb}\}
\end{equation}
where $n_{orb}$ denotes the number of correlated orbitals. And we get the similar constrain for the second Gutzwiller constraint in Eq. (\ref{cst}),
\begin{equation}\label{cst2_n0}
n^B_{i,\alpha} \equiv Tr(\boldsymbol{\phi}^\dag_i \boldsymbol{\phi}_i \boldsymbol{N}_{i\alpha}) =\boldsymbol{a}_i \mathbb{N}_{i\alpha} \boldsymbol{a}^T_i= n^0_{i,\alpha}, \forall \alpha \in \{1, \cdots,n_{orb}\}
\end{equation}
Note that all the $\boldsymbol{n^0}$ occurs in the formulas in previous subsections like Eq. (\ref{m0fock}) and Eq. (\ref{r}) should refer to the variational single particle density matrix.

With the above consideration, we can now obtain the following total energy functional with the constraints embedded through several Lagrange multipliers.
\begin{equation}
\begin{split}
\mathscr{E}[\boldsymbol{n^0},\ket{0},\boldsymbol{a},\boldsymbol{\mathcal{R}}(\boldsymbol{n^0},\boldsymbol{a})] =& \sum_{i\ne j;\alpha,\beta,\delta,\gamma} t^{\alpha\beta}_{ij}\mathcal{R}^\dag_{i,\alpha\gamma} \bra{0}\hat{c}^\dag_{i\gamma} \hat{c}_{j\delta}\ket{0} \mathcal{R}_{j,\delta\beta} \\
&+  \sum_i \boldsymbol{a}_i \mathbb{L}_i \boldsymbol{a}_i^T + E^F(1-\braket{0|0}) \\
&+ \sum_i E^G_i (1-\boldsymbol{a}_i\mathbb{F}_i \boldsymbol{a}_i^\dag) + \sum_{i,\alpha} \lambda^F_{i\alpha}(n^F_{i,\alpha} - n^0_{i,\alpha}) \\
&+  \sum_{i,\alpha} \lambda^B_{i,\alpha}(\boldsymbol{a}_i\mathbb{N}_{i\alpha} \boldsymbol{a}_i^T - n^0_{i,\alpha})
\end{split}
\end{equation}

$E^F$ ($E^G_i$) is the multiplier to ensure the normalization of the single particle wave function (Gutzwiller wave function). Although this procedure has always been implemented automatically by the computer programs, we keep them here for the convenience to build Kohn-Sham like equation. The counterpart of $E^F$ in momentum space of translational invariant systems is the band energy which we are familiar with.  The second term in the fourth line is the constraint explained before. The Lagrange multiplier $\boldsymbol{\lambda^F}$ is tuned to make $\boldsymbol{n^F}$ meet well with the given $\boldsymbol{n^0}$. We will discuss $\boldsymbol{\lambda^F}$ in detail later when introducing the combination of DFT and the Gutzwiller method. The last term is the second Gutzwiller constraint.

For the translational invariant systems which we focus on currently, the site index $i$ of the local quantities which are equivalent on different sites will be dropped for simplicity. For example, the site index $i$ of $\boldsymbol{\phi}_i, \mathbb{L}_i, \mathbb{N}_{i}, \mathbb{F}_{i}, \boldsymbol{n}^0_i, \boldsymbol{n}^F_i, \boldsymbol{\lambda}^F_i, \boldsymbol{\lambda}^B_i, \boldsymbol{a}_i, \boldsymbol{\mathcal{R}}_i$ will be dropped hereafter.

\subsection{Variational scheme}
In this section, we will show in details the variational scheme of this method which consists of two levels of processes: outer loop and inner loop. 

\subsubsection{Outer loop}\label{outloop}
The energy functional of the outermost layer is defined as 
\begin{equation}
\mathscr{E}[\boldsymbol{n^0}] \equiv \mathop{min}_{\ket{0},\boldsymbol{a}} \mathscr{E}[\boldsymbol{n^0},\ket{0},\boldsymbol{a},\boldsymbol{\mathcal{R}}[\boldsymbol{n^0},\boldsymbol{a}]]\Big\vert _{\boldsymbol{n_0}}
\end{equation}
As sketched in Fig. \ref{Fig:il}, we will reach convergence for $\boldsymbol{n^0}$ after we finish the inner loop. The minimization of total energy with respect to $\boldsymbol{n^0}$ will then be performed in outer loop. The optimization of outer loop is constrained by the following relations,
\begin{equation}
\begin{split}
\sum_{\alpha} n^0_\alpha &= N_{corr} \quad \\
 0< &n^0_\alpha < 1  \quad   \forall \alpha \in \{1, \cdots, n_{orb}\}\\
\end{split}
\end{equation}
where $N_{corr}$ is the number of all the correlated electrons.

Up to now, we have successfully transformed the problem to a standard optimization problem with just bound and linear constraints. Besides, the number of variational parameters in outer loop just scales as the product of number of inequivalent atoms and correlated orbitals, which makes it easy to converge.
\begin{figure}[ht]
\includegraphics[width=0.45\textwidth]{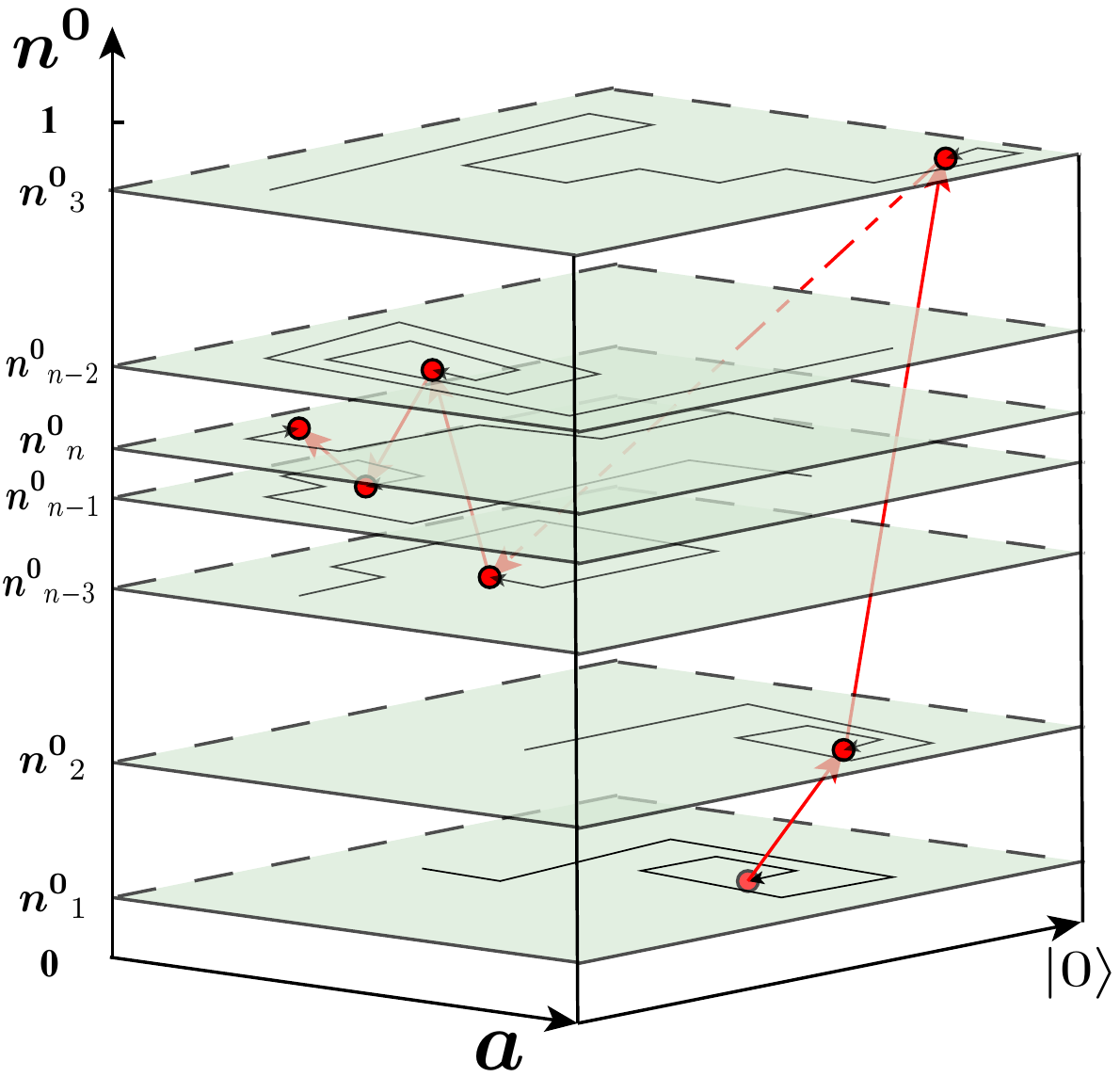}
\centering
\caption{\label{Fig:il}(Color online). Schematic plot of the two-level variational scheme. The light green plane of $\{\ket{0},\boldsymbol{a}\}$ at fixed $\boldsymbol{n^0}$ represents the optimization space of inner loop. The black solid line on each plane means the iteration process when performing minimization of the energy functional with respect to $\{\ket{0},\boldsymbol{a}\}$ for given $\boldsymbol{n^0}$. The red dot in each plane means the corresponding energy minimum. After the inner loop converges, the read line demonstrate the optimization path of energy functional with respect to $\boldsymbol{n^0}$ in outer loop.}
\end{figure}

\subsubsection{Inner loop}\label{sssil}
When $\boldsymbol{n^0}$ has been fixed, the minimization of the total energy should be done with respect to $\ket{0}$ and $\boldsymbol{a}$ in inner loop, which has been sketched in Fig. \ref{fig:flowchart} (a). However, it's hard to perform full optimization in the joint space of $\{\ket{0},\boldsymbol{a}\}$. This problem has been solved by the two-step method proposed by \textit{Deng et al.} \cite{deng2009local}. In the two-step method, $\ket{0}$ is to be solved for given $\boldsymbol{a}$ in so-called "Fermi part" and $\boldsymbol{a}$ is to be determined with $\ket{0}$ fixed in "Bose part", which is shown in Fig. \ref{fig:flowchart} (b). Finally, the complicated minimization can be transferred to the root-searching of a fixed point problem of $\boldsymbol{\mathcal{R}}$ as schematically depicted in Fig. \ref{fig:flowchart} (c) and Fig. \ref{fig:flowchart} (d).

\begin{figure}[ht]
\includegraphics[width=0.45\textwidth]{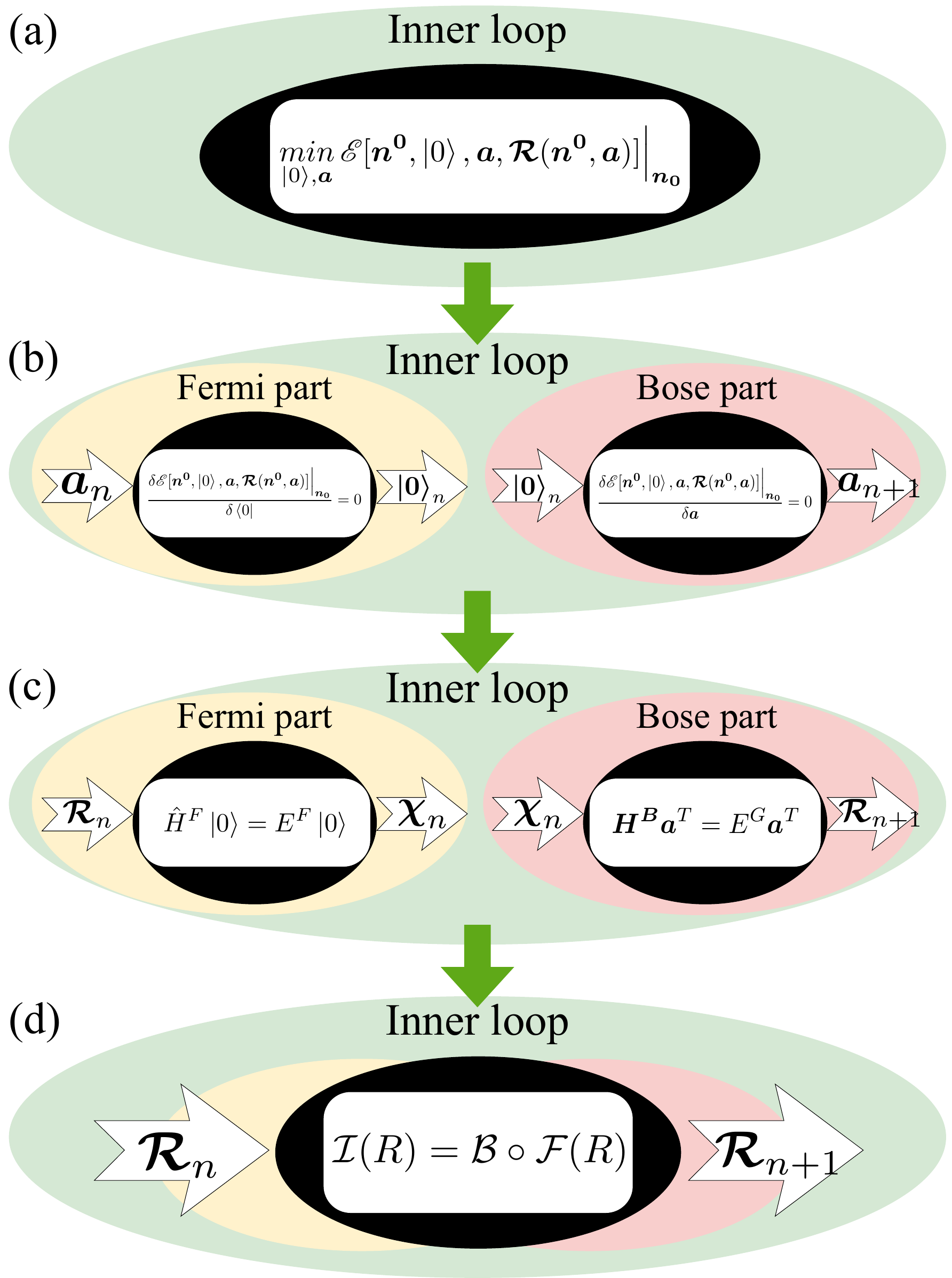}
\centering
\caption{\label{fig:flowchart}(Color online). Schematic plot of the hierarchical construction of the optimization strategy for inner loop. (a) The inner loop originally aims at optimizing $\mathscr{E}[\boldsymbol{n^0},\ket{0},\boldsymbol{a},\boldsymbol{\mathcal{R}}(\boldsymbol{n^0},\boldsymbol{a})]$ with respect to $\ket{0}$ and $\boldsymbol{a}$ on a complex manifold at fixed $\boldsymbol{n^0}$. Take $n$-th step for example. (b) The two parts named the Fermi part and Bose part of inner loop. $\ket{0}_n$ is solved with $\boldsymbol{a}_n$ fixed in Fermi part and $\boldsymbol{a}_{n+1}$ is solved with $\ket{0}_n$ fixed in Bose part. The ellipse filled in black in each subplot is the so-called blackbox. The arrow on the left side of each subplot denotes input while that on the right side means the output of the blackbox. (c) The input and output of both Fermi and Bose parts are equivalently transformed from $\boldsymbol{a}$ and $\ket{0}$ to $\boldsymbol{\chi}$ and $\boldsymbol{\mathcal{R}}$. (d) The Fermi part and Bose part of the inner loop are combined into a united blackbox of which the input is $\boldsymbol{\mathcal{R}}_n$ and the output is $\boldsymbol{\mathcal{R}}_{n+1}$.}
\end{figure}

The energy functional of inner loop at fixed $\boldsymbol{n^0}$ reads, 
\begin{equation}\label{op3}
\begin{split}
\mathscr{E}[\boldsymbol{n^0},\ket{0},\boldsymbol{a},\boldsymbol{\mathcal{R}}(\boldsymbol{n^0},\boldsymbol{a})]\Big\vert _{\boldsymbol{n_0}} =& \sum_{i\ne j;\alpha,\beta,\delta,\gamma} t^{\alpha\beta}_{ij}\mathcal{R}^\dag_{\alpha\gamma} \bra{0}\hat{c}^\dag_{i\gamma} \hat{c}_{j\delta}\ket{0} \mathcal{R}_{\delta\beta} \\
&+  \sum_i \boldsymbol{a} \mathbb{L} \boldsymbol{a}^T\Big\vert _{\boldsymbol{n_0}} + E^F(1-\braket{0|0}) \\
&+ \sum_i E^G (1-\boldsymbol{a}\mathbb{F} \boldsymbol{a}^\dag)\Big\vert _{\boldsymbol{n_0}} \\
&+ \sum_{i,\alpha} \lambda^F_{\alpha}(n^F_{\alpha} - n^0_{\alpha})\Big\vert _{\boldsymbol{n_0}} \\
&+  \sum_{i,\alpha} \lambda^B_{\alpha}(\boldsymbol{a}\mathbb{N}_{\alpha} \boldsymbol{a}^T - n^0_\alpha)\Big\vert _{\boldsymbol{n_0}}
\end{split}
\end{equation}

The search for energy minimum leads to the following stationary equations,

\begin{equation}
\begin{split}\label{se1}
\frac{\delta \mathscr{E}[\boldsymbol{n^0},\ket{0},\boldsymbol{a},\boldsymbol{\mathcal{R}}(\boldsymbol{n^0},\boldsymbol{a})]\Big\vert _{\boldsymbol{n_0}}}{ \delta \bra{0}}  &= 0 \\
\frac{\delta \mathscr{E}[\boldsymbol{n^0},\ket{0},\boldsymbol{a},\boldsymbol{\mathcal{R}}(\boldsymbol{n^0},\boldsymbol{a})]\Big\vert _{\boldsymbol{n_0}}}{ \delta \boldsymbol{a}} &= 0 \quad .
\end{split}
\end{equation}

Substituting Eq. (\ref{op3}) into Eq. (\ref{se1}), the above equations become
\begin{equation}\label{se2}
\begin{split}
\hat{H}^F \ket{0} & = E^F \ket{0} \\
\boldsymbol{H^B} \boldsymbol{a}^T &= E^G \mathbb{F} \boldsymbol{a}^T
\end{split}
\end{equation}
with 
\begin{equation}\label{se3}
\begin{split}
\hat{H}^F(\boldsymbol{n^0},\boldsymbol{\mathcal{R}}(\boldsymbol{n^0},\boldsymbol{a})) &=  \sum_{i\ne j;\alpha,\beta,\delta,\gamma} t^{\alpha\beta}_{ij}\mathcal{R}^\dag_{\alpha\gamma} \hat{c}^\dag_{i\gamma} \hat{c}_{j\delta} \mathcal{R}_{\delta\beta} + \sum_{i,\alpha} \lambda^F_{\alpha} \ket{\alpha}\bra{\alpha}\\
\boldsymbol{H^B}(\boldsymbol{n^0},\boldsymbol{\chi}(\ket{0}))  &= \sum_{\alpha,\beta} \chi_{\alpha\beta} \mathbb{R}_{\alpha\beta} + \sum_{\alpha,\beta} \chi^\dag_{\alpha\beta} \mathbb{R}^\dag_{\alpha\beta} + \mathbb{L} + \sum_\alpha \lambda^B_\alpha \mathbb{N}_\alpha
\end{split}
\end{equation}
where the intermediate parameter $\boldsymbol{\chi}$ is defined as 
\begin{equation}\label{chi}
\chi_{\alpha\beta} = \Big(\frac{\partial E^{kin}}{\partial \mathcal{R}_{\alpha\beta}}\Big)_{\boldsymbol{n^0}} 
\end{equation}
where the subscript $\boldsymbol{n^0}$ means the derivative is done with $\boldsymbol{n^0}$ fixed. $\boldsymbol{\chi}^\dag$ is the Hermitian conjugate of $\boldsymbol{\chi}$.

The effective single particle Hamiltonian depicted in the first one of Eq. (\ref{se3}) is called "Fermi" part while the second one of Eq. (\ref{se3}) called "Bose" part. Fermi part is a Kohn-Sham-like equation which can be diagonalized easily at each $k$ to get the band structure of quasiparticles of Landau Fermi liquid with translational symmetry. The Bose part is a generalized eigenvalue problem in Fock space with matrix $\mathbb{F}$ being positively defined. All the many-body matrices except for  $\mathbb{R}$ are diagonal block between Fock spaces of different particle numbers, while the matrix $\mathbb{R}$ couples the Fock spaces with different particle numbers.

It is difficult to solve the Eq. (\ref{se2}) directly. An alternative way is the two-step method proposed in Ref. \cite{deng2009local}, which solves one of the $\{\ket{0},\boldsymbol{a}\}$ at a time with the other one being fixed and vice versa. As for Fermi part, the Kohn-Sham-like equation depends explicitly on $\ket{0}$ and implicitly on $\boldsymbol{a}$ through the orbital renormalization matrix $\boldsymbol{\mathcal{R}}(\boldsymbol{n^0},\boldsymbol{a})$. Therefore, we will obtain $\ket{0}$ by diagonalizing the hamiltonian with $\boldsymbol{a}$ being fixed, or equally speaking,  with $\boldsymbol{\mathcal{R}}(\boldsymbol{n^0},\boldsymbol{a})$ being fixed. As for the Bose part, the eigenvalue problem depends explicitly on $\boldsymbol{a}$ and implicitly on $\ket{0}$ through the intermediate parameter $\boldsymbol{\chi}(\ket{0})$. Hence $\boldsymbol{a}$ is calculated by diagonalizing the many-body Hamiltonian with $\ket{0}$ being fixed, or equally speaking, with $\boldsymbol{\chi}(\ket{0})$ being fixed. As being sketched in Fig. \ref{fig:flowchart} (b) and (c), the Fermi part colored in light yellow sends out the intermediate parameter $\boldsymbol{\chi}$ as the output after receiving $\boldsymbol{\mathcal{R}}$ as input. On the contrary, the Bose part in light pink takes $\boldsymbol{\chi}$ as the input which is the output of Fermi part and compute the matrix $\boldsymbol{\mathcal{R}}$. The Fermi part and Bose part can be combined into a united iteration loop as shown in Fig. \ref{fig:flowchart}(d). The corresponding composite function can be written as  

\begin{equation}
\mathcal{I}(\boldsymbol{\mathcal{R}}) = \mathcal{B} \circ \mathcal{F}(\boldsymbol{\mathcal{R}})
\end{equation}
with
\begin{equation}\label{fb2}
\begin{split}
\mathcal{F}(\boldsymbol{\mathcal{R}_n}) & = \boldsymbol{\chi_n} \\
\mathcal{B}(\boldsymbol{\chi_n}) &= \boldsymbol{\mathcal{R}_{n+1}} 
\end{split}
\end{equation}
where $\mathcal{F}$ represents the Fermi part and $\mathcal{B}$ the Bose part.

 The single particle part of the Hamiltonian can be always Fourier transformed into the momentum space for convenience. With the definition of Bloch basis $\ket{\boldsymbol{k}\alpha} = {1 \over \sqrt{N_{R}}} \sum_i e^{i\boldsymbol{k}\cdot\boldsymbol{R}_i} \ket{i,\alpha}$, the effective single particle Hamiltonian of the Fermi part is expressed as 
 
 \begin{equation}
 \hat{H}^F_{\boldsymbol{k}} = \sum_{\alpha,\beta,\delta,\gamma} t^{\alpha\beta}_{\boldsymbol{k}}\mathcal{R}^\dag_{\alpha\gamma} \hat{c}^\dag_{\boldsymbol{k}\gamma} \hat{c}_{\boldsymbol{k}\delta} \mathcal{R}_{\delta\beta} + \sum_\alpha \lambda^F_\alpha \ket{\boldsymbol{k}\alpha}\bra{\boldsymbol{k}\alpha}
 \end{equation}

Note that the Lagrange multipliers $\boldsymbol{\lambda^F}$ and $\boldsymbol{\lambda^B}$ are yet to be determined. In Fermi part, $\boldsymbol{\lambda^F}$ will be tuned to fulfill the constraint in Eq. (\ref{cst1_n0}). And $\boldsymbol{\lambda^B}$ of Bose part will be adjusted to satisfy the constraint in Eq. (\ref{cst2_n0}). How to search $\boldsymbol{\lambda^F}$ and $\boldsymbol{\lambda^B}$ efficiently will be introduced in the next subsection.

\subsection{Optimization}\label{opt}
Both the outer and inner loops numerically reply on the specific implementation of optimization and root search. 
In this section, we will demonstrate in details how to optimize the energy functional with respect to $\boldsymbol{n^0}$ in the outer loop and search the root of the fixed point problem in the inner loop using the information of the derivatives. In order to reach the stable self consistent solution for various situations, we need to use derivatives in both the outer loop minimization and inner loop root search. We will then introduce the equations to obtain both the full and partial derivatives in this section.

Firstly, let's distinguish the full derivative (FD) and partial derivative (PD) as well as partial partial derivative (PPD). When we talk about the FD of an operator named as $\hat{A}$, it means the change rate of $\hat{A}$ with respect to the infinitesimal change of $\boldsymbol{n^0}$ after the inner loop reaches new convergence. In the Fermi part, the PD of $\boldsymbol{\chi}$ means the changing rate of $\boldsymbol{\chi}$ respect to the infinitesimal change of one of the $\{\boldsymbol{n^0},\boldsymbol{\mathcal{R}}\}$ with the other one being fixed. In the Bose part, the PD of $\boldsymbol{\mathcal{R}}$ refers to the changing rate of $\boldsymbol{\mathcal{R}}$ respect to one of the $\{\boldsymbol{n^0},\boldsymbol{\chi}\}$ with the other one being fixed. Please refer to Fig. \ref{fig:flowchart} for illustration. The final one is PPD which is similar as PD but taking $\{\boldsymbol{n^0},\boldsymbol{\mathcal{R}},\boldsymbol{\lambda^F}\}$ as independent variables for Fermi part while $\{\boldsymbol{n^0},\boldsymbol{\chi},\boldsymbol{\lambda^B}\}$ for Bose part. Specifically, take Fermi part for example, PPD of $\boldsymbol{\chi}$ with respect to $\boldsymbol{\lambda^F}$ means the changing rate of $\boldsymbol{\chi}$ relative to the infinitesimal change of  $\boldsymbol{\lambda^F}$ with $\boldsymbol{n^0}$ and $\boldsymbol{\mathcal{R}}$ being fixed.  It is obvious that all the derivatives of the DFT+G method are "process derivative" which means the derivative of the practical implementation process. By applying the chain rule, we can obtain the PDs from the PPDs and FDs from the PDs. Note that $\frac{d (\cdots)}{d (\cdots)} $ denotes FD, $\frac{\partial (\cdots)}{\partial (\cdots)} $ means PD and $\frac{\partial_0 (\cdots)}{\partial_0 (\cdots)}$ represents PPD in the subsequent formulas. In order to distinguish the variables in different iteration steps, they will be labelled number as subscript to indicate which iteration the variables belong to. For conciseness, the subscript will be omitted when the variables are at the $n$-th step which is the default set of the current step.

\subsubsection{FDs}\label{fds}

\begin{figure}[ht]
\includegraphics[width=0.49\textwidth]{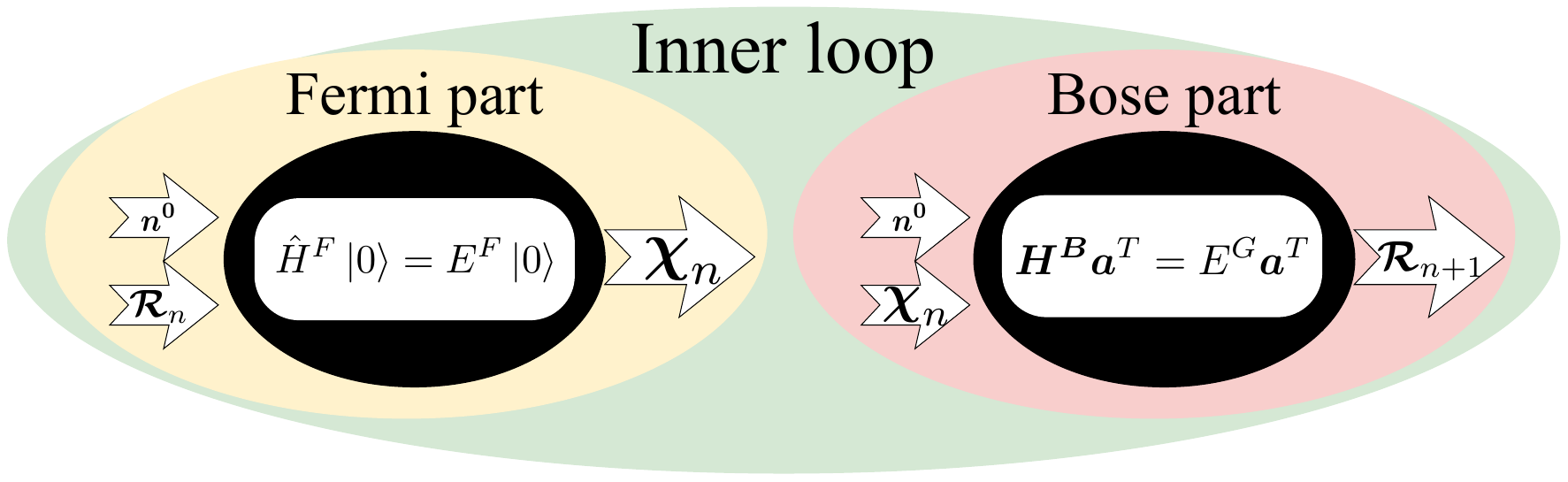}
\centering
\caption{\label{Fig:fb4}(Color online). Schematic plot for the Fermi part and Bose part at $n$-th step with $\boldsymbol{n^0}$ not fixed. For Fermi part in light yellow, the output $\boldsymbol{\chi}_n$ varies with $\boldsymbol{\mathcal{R}}_n$ and $\boldsymbol{n^0}$. For Bose part in light pink, the output $\boldsymbol{\mathcal{R}}_{n+1}$ varies with $\boldsymbol{\chi}_n$ and $\boldsymbol{n^0}$.}
\end{figure}

The total energy including kinetic and interaction energy is shown in Eq. (\ref{op1}). Then the global derivatives of total energy with respect of $\boldsymbol{n^0}$ can be written formally as

\begin{widetext}
\begin{equation}\label{td}
\begin{split}
\frac{d \mathscr{E}[\boldsymbol{n^0}]}{d \boldsymbol{n^0}}  =& \frac{d E^{kin}[\boldsymbol{n^0}]}{d \boldsymbol{n^0}} + \frac{d E^{a}[\boldsymbol{n^0}]}{d \boldsymbol{n^0}} \\
=& \frac{\partial E^{kin}}{\partial \boldsymbol{n^0}} +  \frac{\partial E^{kin}}{\partial \boldsymbol{\mathcal{R}}} \frac{d \boldsymbol{\mathcal{R}}}{d \boldsymbol{n^0}} +  \frac{\partial E^{kin}}{\partial \boldsymbol{\mathcal{R}^\dag}} \frac{d \boldsymbol{\mathcal{R}}^\dag}{d \boldsymbol{n^0}} 
+ \frac{\partial E^{a}}{\partial \boldsymbol{n^0}} + \frac{\partial E^a}{\partial \boldsymbol{\chi}} \frac{d \boldsymbol{\chi}}{d \boldsymbol{n^0}} + \frac{\partial E^a}{\partial \boldsymbol{\chi}^\dag} \frac{d \boldsymbol{\chi}^\dag}{d \boldsymbol{n^0}}\\
=& \frac{\partial E^{kin}}{\partial \boldsymbol{n^0}} + \boldsymbol{\chi} \frac{d \boldsymbol{\mathcal{R}}}{d \boldsymbol{n^0}} + \boldsymbol{\chi}^\dag \frac{d \boldsymbol{\mathcal{R}}^\dag}{d \boldsymbol{n^0}} 
+ \frac{\partial E^{a}}{\partial \boldsymbol{n^0}} + \frac{\partial E^a}{\partial \boldsymbol{\chi}} \frac{d \boldsymbol{\chi}}{d \boldsymbol{n^0}} + \frac{\partial E^a}{\partial \boldsymbol{\chi}^\dag} \frac{d \boldsymbol{\chi}^\dag}{d \boldsymbol{n^0}}
\end{split}
\end{equation} 
\end{widetext}

There are four total derivatives remaining in Eq. (\ref{td}) need to be calculated. Using the linear response theory for Fermi part and referring to Fig. (\ref{Fig:fb4}), it's easy to develop
\begin{equation}\label{lr1}
\begin{cases}
\delta\boldsymbol{\chi} &= \frac{\partial \boldsymbol{\chi}}{\partial \boldsymbol{n^0}}  \delta \boldsymbol{n^0}  + \frac{\partial \boldsymbol{\chi}}{\partial \boldsymbol{\mathcal{R}}}\delta \boldsymbol{\mathcal{R}} + \frac{\partial \boldsymbol{\chi}}{\partial \boldsymbol{\mathcal{R}}^\dag} \delta \boldsymbol{\mathcal{R}}^\dag\\
\delta\boldsymbol{\chi}^\dag &= \frac{\partial \boldsymbol{\chi}^\dag}{\partial \boldsymbol{n^0}}  \delta \boldsymbol{n^0}  + \frac{\partial \boldsymbol{\chi}^\dag}{\partial \boldsymbol{\mathcal{R}}}\delta \boldsymbol{\mathcal{R}} + \frac{\partial \boldsymbol{\chi}^\dag}{\partial \boldsymbol{\mathcal{R}}^\dag} \delta \boldsymbol{\mathcal{R}}^\dag
\end{cases}
\end{equation}
where "$\delta$" in the right hand side denotes infinitesimal change of independent variables while that in the left hand side means the response of the dependent variables. Notice here the $\boldsymbol{n^0}$ is independent variables not just a fixed parameter when we consider the full derivatives from the top most layer.

Similar relations are also derived for Bose part,
\begin{equation}\label{lr2}
\begin{cases}
\delta \boldsymbol{\mathcal{R}}_{n+1} &= \frac{\partial \boldsymbol{\mathcal{R}}_{n+1}}{\partial \boldsymbol{n^0}}  \delta \boldsymbol{n^0}  + \frac{\partial \boldsymbol{\mathcal{R}}_{n+1}}{\partial \boldsymbol{\chi}}\delta \boldsymbol{\chi} + \frac{\partial \boldsymbol{\mathcal{R}}_{n+1}}{\partial \boldsymbol{\chi}^\dag} \delta \boldsymbol{\chi}^\dag\\
\delta \boldsymbol{\mathcal{R}}^\dag_{n+1} &= \frac{\partial \boldsymbol{\mathcal{R}}^\dag_{n+1}}{\partial \boldsymbol{n^0}}  \delta \boldsymbol{n^0}  + \frac{\partial \boldsymbol{\mathcal{R}}^\dag_{n+1}}{\partial \boldsymbol{\chi}}\delta \boldsymbol{\chi} + \frac{\partial \boldsymbol{\mathcal{R}}^\dag_{n+1}}{\partial \boldsymbol{\chi}^\dag} \delta \boldsymbol{\chi}^\dag
\end{cases}
\end{equation}

We obtain the following equation through elimination after substituting Eq. (\ref{lr1}) into Eq. (\ref{lr2}), 
\begin{widetext}
\begin{equation}
\delta \boldsymbol{\mathcal{R}} = (1-\boldsymbol{C})^{-1} (\boldsymbol{A}+\boldsymbol{B}) \delta \boldsymbol{n^0} 
\end{equation}
with 
\begin{equation}\label{l1}
\begin{split}
\boldsymbol{A}&=(1 -  \frac{\partial \boldsymbol{\mathcal{R}}_{n+1}}{\partial \boldsymbol{\chi}}\frac{\partial \boldsymbol{\chi}}{\partial \boldsymbol{\mathcal{R}}} -  \frac{\partial \boldsymbol{\mathcal{R}}_{n+1}}{\partial \boldsymbol{\chi}^\dag}\frac{\partial \boldsymbol{\chi}^\dag}{\partial \boldsymbol{\mathcal{R}}} )^{-1}(\frac{\partial \boldsymbol{\mathcal{R}}_{n+1}}{\partial \boldsymbol{n^0}} +  \frac{\partial \boldsymbol{\mathcal{R}}_{n+1}}{\partial \boldsymbol{\chi}} \frac{\partial \boldsymbol{\chi}}{\partial \boldsymbol{n^0}}  + \frac{\partial \boldsymbol{\mathcal{R}}_{n+1}}{\partial \boldsymbol{\chi}^\dag} \frac{\partial \boldsymbol{\chi}^\dag}{\partial \boldsymbol{n^0}} )\\
\boldsymbol{B}&=(1 -  \frac{\partial \boldsymbol{\mathcal{R}}_{n+1}}{\partial \boldsymbol{\chi}}\frac{\partial \boldsymbol{\chi}}{\partial \boldsymbol{\mathcal{R}}} -  \frac{\partial \boldsymbol{\mathcal{R}}_{n+1}}{\partial \boldsymbol{\chi}^\dag}\frac{\partial \boldsymbol{\chi}^\dag}{\partial \boldsymbol{\mathcal{R}}} )^{-1}( \frac{\partial \boldsymbol{\mathcal{R}}_{n+1}}{\partial \boldsymbol{\chi}} \frac{\partial \boldsymbol{\chi}}{\partial \boldsymbol{\mathcal{R}}^\dag} +  \frac{\partial \boldsymbol{\mathcal{R}}_{n+1}}{\partial \boldsymbol{\chi}^\dag}  \frac{\partial \boldsymbol{\chi}^\dag}{\partial \boldsymbol{\mathcal{R}}^\dag})(1 -  \frac{\partial \boldsymbol{\mathcal{R}}^\dag_{n+1}}{\partial \boldsymbol{\chi}}\frac{\partial \boldsymbol{\chi}}{\partial \boldsymbol{\mathcal{R}}^\dag} -  \frac{\partial \boldsymbol{\mathcal{R}}^\dag_{n+1}}{\partial \boldsymbol{\chi}^\dag}\frac{\partial \boldsymbol{\chi}^\dag}{\partial \boldsymbol{\mathcal{R}}^\dag} )^{-1}(\frac{\partial \boldsymbol{\mathcal{R}}^\dag_{n+1}}{\partial \boldsymbol{n^0}} +  \frac{\partial \boldsymbol{\mathcal{R}}^\dag_{n+1}}{\partial \boldsymbol{\chi}} \frac{\partial \boldsymbol{\chi}}{\partial \boldsymbol{n^0}}  + \frac{\partial \boldsymbol{\mathcal{R}}^\dag_{n+1}}{\partial \boldsymbol{\chi}^\dag} \frac{\partial \boldsymbol{\chi}^\dag}{\partial \boldsymbol{n^0}} )\\
\boldsymbol{C}&=(1 -  \frac{\partial \boldsymbol{\mathcal{R}}_{n+1}}{\partial \boldsymbol{\chi}}\frac{\partial \boldsymbol{\chi}}{\partial \boldsymbol{\mathcal{R}}} -  \frac{\partial \boldsymbol{\mathcal{R}}_{n+1}}{\partial \boldsymbol{\chi}^\dag}\frac{\partial \boldsymbol{\chi}^\dag}{\partial \boldsymbol{\mathcal{R}}} )^{-1}( \frac{\partial \boldsymbol{\mathcal{R}}_{n+1}}{\partial \boldsymbol{\chi}} \frac{\partial\boldsymbol{\chi}}{\partial \boldsymbol{\mathcal{R}}^\dag} +  \frac{\partial \boldsymbol{\mathcal{R}}_{n+1}}{\partial \boldsymbol{\chi}^\dag}  \frac{\partial \boldsymbol{\chi}^\dag}{\partial \boldsymbol{\mathcal{R}}^\dag})(1 -  \frac{\partial \boldsymbol{\mathcal{R}}^\dag_{n+1}}{\partial \boldsymbol{\chi}}\frac{\partial \boldsymbol{\chi}}{\partial \boldsymbol{\mathcal{R}}^\dag} -  \frac{\partial \boldsymbol{\mathcal{R}}^\dag_{n+1}}{\partial\boldsymbol{\chi}^\dag}\frac{\partial \boldsymbol{\chi}^\dag}{\partial \boldsymbol{\mathcal{R}}^\dag} )^{-1}( \frac{\partial \boldsymbol{\mathcal{R}}^\dag_{n+1}}{\partial \boldsymbol{\chi}} \frac{\partial \boldsymbol{\chi}}{\partial \boldsymbol{\mathcal{R}}} +  \frac{\partial \boldsymbol{\mathcal{R}}^\dag_{n+1}}{\partial \boldsymbol{\chi}^\dag}  \frac{\partial \boldsymbol{\chi}^\dag}{\partial \boldsymbol{\mathcal{R}}})\\
\end{split}
\end{equation}

\newpage

After substituting Eq. (\ref{lr2}) into Eq. (\ref{lr1}), the similar response equation for $\boldsymbol{\chi}$ reads,
\begin{equation}
\delta \boldsymbol{\chi} = (1-\boldsymbol{C}')^{-1} (\boldsymbol{A}'+\boldsymbol{B}') \delta \boldsymbol{n^0} 
\end{equation}
with

\begin{equation}\label{l2}
\begin{split}
\boldsymbol{A}'&=(1 - \frac{\partial \boldsymbol{\chi}}{\partial \boldsymbol{\mathcal{R}}} \frac{\partial \boldsymbol{\mathcal{R}}_{n+1}}{\partial \boldsymbol{\chi}} -  \frac{\partial \boldsymbol{\chi}}{\partial \boldsymbol{\mathcal{R}}^\dag}\frac{\partial \boldsymbol{\mathcal{R}}^\dag_{n+1}}{\partial \boldsymbol{\chi}} )^{-1}(\frac{\partial \boldsymbol{\chi}}{\partial \boldsymbol{n^0}} +  \frac{\partial \boldsymbol{\chi}}{\partial \boldsymbol{\mathcal{R}}} \frac{\partial \boldsymbol{\mathcal{R}}_{n+1}}{\partial \boldsymbol{n^0}}  + \frac{\partial \boldsymbol{\chi}}{\partial \boldsymbol{\mathcal{R}}^\dag} \frac{\partial \boldsymbol{\mathcal{R}}^\dag_{n+1}}{\partial \boldsymbol{n^0}} )\\
\boldsymbol{B}'&=(1 - \frac{\partial \boldsymbol{\chi}}{\partial \boldsymbol{\mathcal{R}}} \frac{\partial \boldsymbol{\mathcal{R}}_{n+1}}{\partial \boldsymbol{\chi}} -  \frac{\partial \boldsymbol{\chi}}{\partial \boldsymbol{\mathcal{R}}^\dag}\frac{\partial \boldsymbol{\mathcal{R}}^\dag_{n+1}}{\partial \boldsymbol{\chi}} )^{-1}( \frac{\partial \boldsymbol{\chi}}{\partial \boldsymbol{\mathcal{R}}} \frac{\partial \boldsymbol{\mathcal{R}}_{n+1}}{\partial \boldsymbol{\chi}^\dag} +  \frac{\partial \boldsymbol{\chi}}{\partial \boldsymbol{\mathcal{R}}^\dag}  \frac{\partial \boldsymbol{\mathcal{R}}^\dag_{n+1}}{\partial \boldsymbol{\chi}^\dag})(1 -  \frac{\partial \boldsymbol{\chi}^\dag}{\partial \boldsymbol{\mathcal{R}}}\frac{\partial \boldsymbol{\mathcal{R}}_{n+1}}{\partial \boldsymbol{\chi}^\dag} -  \frac{\partial \boldsymbol{\chi}^\dag}{\partial \boldsymbol{\mathcal{R}}^\dag}\frac{\partial \boldsymbol{\mathcal{R}}^\dag_{n+1}}{\partial \boldsymbol{\chi}^\dag} )^{-1}(\frac{\partial \boldsymbol{\chi}^\dag}{\partial \boldsymbol{n^0}} +  \frac{\partial \boldsymbol{\chi}^\dag}{\partial \boldsymbol{\mathcal{R}}} \frac{\partial \boldsymbol{\mathcal{R}}_{n+1}}{\partial \boldsymbol{n^0}}  + \frac{\partial \boldsymbol{\chi}^\dag}{\partial \boldsymbol{\mathcal{R}}^\dag} \frac{\partial \boldsymbol{\mathcal{R}}^\dag_{n+1}}{\partial \boldsymbol{n^0}} )\\
\boldsymbol{C}'&=(1 - \frac{\partial \boldsymbol{\chi}}{\partial \boldsymbol{\mathcal{R}}} \frac{\partial \boldsymbol{\mathcal{R}}_{n+1}}{\partial \boldsymbol{\chi}} -  \frac{\partial \boldsymbol{\chi}}{\partial \boldsymbol{\mathcal{R}}^\dag}\frac{\partial \boldsymbol{\mathcal{R}}^\dag_{n+1}}{\partial \boldsymbol{\chi}} )^{-1}( \frac{\partial \boldsymbol{\chi}}{\partial \boldsymbol{\mathcal{R}}} \frac{\partial \boldsymbol{\mathcal{R}}_{n+1}}{\partial \boldsymbol{\chi}^\dag} +  \frac{\partial \boldsymbol{\chi}}{\partial \boldsymbol{\mathcal{R}}^\dag}  \frac{\partial \boldsymbol{\mathcal{R}}^\dag_{n+1}}{\partial \boldsymbol{\chi}^\dag})(1 -  \frac{\partial \boldsymbol{\chi}^\dag}{\partial \boldsymbol{\mathcal{R}}}\frac{\partial \boldsymbol{\mathcal{R}}_{n+1}}{\partial \boldsymbol{\chi}^\dag} -  \frac{\partial \boldsymbol{\chi}^\dag}{\partial \boldsymbol{\mathcal{R}}^\dag}\frac{\partial \boldsymbol{\mathcal{R}}^\dag_{n+1}}{\partial \boldsymbol{\chi}^\dag} )^{-1}( \frac{\partial \boldsymbol{\chi}^\dag}{\partial \boldsymbol{\mathcal{R}}} \frac{\partial \boldsymbol{\mathcal{R}}_{n+1}}{\partial \boldsymbol{\chi}} +  \frac{\partial \boldsymbol{\chi}^\dag}{\partial \boldsymbol{\mathcal{R}}^\dag}  \frac{\partial \boldsymbol{\mathcal{R}}^\dag_{n+1}}{\partial \boldsymbol{\chi}})\\
\end{split}
\end{equation}
\end{widetext}

\newpage
Thus the full derivatives of both $\boldsymbol{\mathcal{R}}$ and $\boldsymbol{\chi}$ with respect to $\boldsymbol{n^0}$ is established as 
\begin{equation}
\begin{split}
\frac{d \boldsymbol{\mathcal{R}}}{d \boldsymbol{n^0}} &= (1-\boldsymbol{C})^{-1}(\boldsymbol{A}+\boldsymbol{B}) \\
\frac{d \boldsymbol{\chi}}{d \boldsymbol{n^0}} &=  (1-\boldsymbol{C}')^{-1}(\boldsymbol{A}'+\boldsymbol{B}')
\end{split}
\end{equation}

Because $\delta \boldsymbol{n^0}$ is a real vector,  we have $\frac{d \mathcal{R}^\dag_{\alpha\beta}}{d \boldsymbol{n^0}}=\Big(\frac{d \mathcal{R}_{\beta\alpha}}{d \boldsymbol{n^0}}\Big)^*$ and $\frac{d \chi^\dag_{\alpha\beta}}{d \boldsymbol{n^0}}=\Big(\frac{d \chi_{\beta\alpha}}{d \boldsymbol{n^0}}\Big)^*$.

\subsubsection{PDs and PPDs}\label{pds}
In this section, we will derive PDs using PPDs through the chain rule. The derivation of PPDs is always based on the second order perturbation theory and will be put in the appendix. $\boldsymbol{\chi}$, $\frac{\partial \boldsymbol{\chi}}{\partial \boldsymbol{n^0}}$, $\frac{\partial \boldsymbol{\chi}}{\partial \boldsymbol{\mathcal{R}}}$, $\frac{\partial \boldsymbol{\chi}^\dag}{\partial \boldsymbol{\mathcal{R}}}$ as well as $\frac{\partial E^{kin}}{\partial \boldsymbol{n^0}}$ in Fermi part and $\frac{\partial \boldsymbol{\mathcal{R}}_{n+1}}{\partial \boldsymbol{n^0}}$, $\frac{\partial \boldsymbol{\mathcal{R}}_{n+1}}{\partial \boldsymbol{\chi}}$, $\frac{\partial \boldsymbol{\mathcal{R}}_{n+1}}{\partial \boldsymbol{\chi}^\dag}$, $\frac{\partial E^a}{\partial \boldsymbol{n^0}}$ as well as $\frac{\partial E^a}{\partial \boldsymbol{\chi}}$ in Bose part are to be derived. For later convenience, $\frac{\partial \boldsymbol{\mathcal{R}}_{n+1}}{\partial \boldsymbol{\mathcal{R}}}$ for the whole inner loop will be also derived.

\begin{widetext}
PDs in Fermi part,
\begin{itemize}
\item $\boldsymbol{\chi}$
\begin{equation}
\boldsymbol{\chi} = \frac{\partial E^{kin}}{\partial \boldsymbol{\mathcal{R}}} = \frac{\partial_0 E^{kin}}{\partial_0 \boldsymbol{\mathcal{R}}}+ \frac{\partial_0 E^{kin}}{\partial_0 \boldsymbol{\lambda^F}}\frac{\partial \boldsymbol{\lambda^F}}{\partial \boldsymbol{\mathcal{R}}} = \frac{\partial_0 E^{kin}}{\partial_0 \boldsymbol{\mathcal{R}}}- \frac{\partial_0 E^{kin}}{\partial_0 \boldsymbol{\lambda^F}}(\frac{\partial_0 \boldsymbol{n^F}}{\partial_0 \boldsymbol{\lambda^F}})^{-1}\frac{\partial_0 \boldsymbol{n^F}}{\partial_0 \boldsymbol{\mathcal{R}}}\\
\end{equation}
\item $\frac{\partial \boldsymbol{\chi}}{\partial \boldsymbol{n^0}}$
\begin{equation}
\frac{\partial \boldsymbol{\chi}}{\partial \boldsymbol{n^0}} =\frac{\partial_0 \boldsymbol{\chi}}{\partial_0 \boldsymbol{\lambda^F}}\frac{\partial \boldsymbol{\lambda^F}}{\partial \boldsymbol{n^0}} = \frac{\partial_0 \boldsymbol{\chi}}{\partial_0 \boldsymbol{\lambda^F}}(\frac{\partial_0 \boldsymbol{n^F}}{\partial_0 \boldsymbol{\lambda^F}})^{-1}
\end{equation}

\item $\frac{\partial \boldsymbol{\chi}}{\partial \boldsymbol{\mathcal{R}}}$
\begin{equation}
\frac{\partial \boldsymbol{\chi}}{\partial \boldsymbol{\mathcal{R}}} = \frac{\partial_0 \boldsymbol{\chi}}{\partial_0 \boldsymbol{\mathcal{R}}} +  \frac{\partial_0 \boldsymbol{\chi}}{\partial_0 \boldsymbol{\lambda^F}}\frac{\partial \boldsymbol{\lambda^F}}{\partial \boldsymbol{\mathcal{R}}}=\frac{\partial_0 \boldsymbol{\chi}}{\partial_0 \boldsymbol{\mathcal{R}}}  -  \frac{\partial_0 \boldsymbol{\chi}}{\partial_0 \boldsymbol{\lambda^F}}(\frac{\partial_0 \boldsymbol{n^F}}{\partial_0 \boldsymbol{\lambda^F}})^{-1}\frac{\partial_0 \boldsymbol{n^F}}{\partial_0 \boldsymbol{\mathcal{R}}}
\end{equation}
And $\frac{\partial \boldsymbol{\chi}^\dag}{\partial \boldsymbol{\mathcal{R}}}$ can be derived similarly.

\item $\frac{\partial E^{kin}}{\partial \boldsymbol{n^0}}$
\begin{equation}
\frac{\partial E^{kin}}{\partial \boldsymbol{n^0}}  = \frac{\partial_0 E^{kin}}{\partial_0 \boldsymbol{\lambda^F} }\frac{\partial \boldsymbol{\lambda^F}}{\partial \boldsymbol{n^0}} = \frac{\partial_0 E^{kin}}{\partial_0 \boldsymbol{\lambda^F}}(\frac{\partial_0 \boldsymbol{n^F}}{\partial_0 \boldsymbol{\lambda^F}})^{-1}
\end{equation}
\end{itemize}

PDs in Bose part,
\begin{itemize}
\item $\frac{\partial \boldsymbol{\mathcal{R}}_{n+1}}{\partial \boldsymbol{n^0}}$ 
\begin{equation}
\frac{\partial \boldsymbol{\mathcal{R}}_{n+1}}{\partial \boldsymbol{n^0}} = \frac{\partial_0 \boldsymbol{\mathcal{R}}_{n+1}}{\partial_0 \boldsymbol{n^0}} + \frac{\partial_0 \boldsymbol{\mathcal{R}}_{n+1}}{\partial_0 \boldsymbol{\lambda^B}} \frac{\partial \boldsymbol{\lambda^B}}{\partial \boldsymbol{n^0}} =  \frac{\partial_0 \boldsymbol{\mathcal{R}}_{n+1}}{\partial_0 \boldsymbol{n^0}} + \frac{\partial_0 \boldsymbol{\mathcal{R}}_{n+1}}{\partial_0 \boldsymbol{\lambda^B}} (\frac{\partial_0 \boldsymbol{n^B}}{\partial_0 \boldsymbol{\lambda^B}})^{-1}(1-\frac{\partial_0 \boldsymbol{n^B}}{\partial_0 \boldsymbol{n^0}})
\end{equation}

\item $\frac{\partial \boldsymbol{\mathcal{R}}_{n+1}}{\partial \boldsymbol{\chi}}$
\begin{equation}
\frac{\partial \boldsymbol{\mathcal{R}}_{n+1}}{\partial \boldsymbol{\chi}} = \frac{\partial_0 \boldsymbol{\mathcal{R}}_{n+1}}{\partial_0 \boldsymbol{\chi}} + \frac{\partial_0 \boldsymbol{\mathcal{R}}_{n+1}}{\partial_0 \boldsymbol{\lambda^B}}\frac{\partial \boldsymbol{\lambda^B}}{\partial \boldsymbol{\chi}} = \frac{\partial_0 \boldsymbol{\mathcal{R}}_{n+1}}{\partial_0 \boldsymbol{\chi}} - \frac{\partial_0 \boldsymbol{\mathcal{R}}_{n+1}}{\partial_0 \boldsymbol{\lambda^B}}(\frac{\partial_0 \boldsymbol{n^B}}{\partial_0 \boldsymbol{\lambda^B}})^{-1}\frac{\partial_0 \boldsymbol{n^B}}{\partial_0 \boldsymbol{\chi}}
\end{equation}
And $\frac{\partial \boldsymbol{\mathcal{R}}_{n+1}}{\partial \boldsymbol{\chi}^\dag}$ can also be derived similarly.

\item $\frac{\partial E^a}{\partial \boldsymbol{n^0}}$
\begin{equation}
\frac{\partial E^a}{\partial \boldsymbol{n^0}}= \frac{\partial_0 E^a}{\partial_0 \boldsymbol{n^0}} + \frac{\partial_0 E^a}{\partial_0 \boldsymbol{\lambda^B}}\frac{\partial \boldsymbol{\lambda^B}}{\partial \boldsymbol{n^0}} =  \frac{\partial_0 E^a}{\partial_0 \boldsymbol{n^0}} + \frac{\partial_0 E^a}{\partial_0 \boldsymbol{\lambda^B}}(\frac{\partial_0 \boldsymbol{n^B}}{\partial_0 \boldsymbol{\lambda^B}})^{-1}(1-\frac{\partial_0 \boldsymbol{n^B}}{\partial_0 \boldsymbol{n^0}})
\end{equation}
\item $\frac{\partial E^a}{\partial \boldsymbol{\chi}}$
\begin{equation}
\frac{\partial E^a}{\partial \boldsymbol{\chi}} = \frac{\partial_0 E^a}{\partial_0 \boldsymbol{\chi}} + \frac{\partial_0 E^a}{\partial_0 \boldsymbol{\lambda^B}}\frac{\partial \boldsymbol{\lambda^B}}{\partial \boldsymbol{\chi}} = \frac{\partial_0 E^a}{\partial_0 \boldsymbol{\chi}}  -\frac{\partial_0 E^a}{\partial_0 \boldsymbol{\lambda^B}}(\frac{\partial_0 \boldsymbol{n^B}}{\partial_0 \boldsymbol{\lambda^B}})^{-1}\frac{\partial_0 \boldsymbol{n^B}}{\partial_0 \boldsymbol{\chi}}
\end{equation}
\end{itemize}

PDs in inner loop,
\begin{itemize}
\item $\frac{\partial \boldsymbol{\mathcal{R}}_{n+1}}{\partial \boldsymbol{\mathcal{R}}}$
\begin{equation}\label{prpr}
\frac{\partial \boldsymbol{\mathcal{R}}_{n+1}}{\partial \boldsymbol{\mathcal{R}}} = \frac{\partial \boldsymbol{\mathcal{R}}_{n+1}}{\partial \boldsymbol{\chi}}\frac{\partial \boldsymbol{\chi}}{\partial \boldsymbol{\mathcal{R}}} + \frac{\partial \boldsymbol{\mathcal{R}}_{n+1}}{\partial \boldsymbol{\chi}^\dag}\frac{\partial \boldsymbol{\chi}^\dag}{\partial \boldsymbol{\mathcal{R}}}
\end{equation}
From the definition, it is obvious that $\frac{\partial \boldsymbol{\mathcal{R}}_{n+1}}{\partial \boldsymbol{\mathcal{R}}}$ is a typical "process derivative". For example, $\frac{\partial \boldsymbol{\mathcal{R}}_{n+1}}{\partial \boldsymbol{\mathcal{R}}_n}\Big\vert_{\boldsymbol{\mathcal{R}}_n}$measure the change rate of the output of inner part ($\boldsymbol{\mathcal{R}}_{n+1} = \mathcal{I}(\boldsymbol{\mathcal{R}}_n)$) with respect to the infinitesimal change of $\boldsymbol{\mathcal{R}}_n$ which is the input of inner part for $n$-th step.

\end{itemize}
\end{widetext}

\subsubsection{Optimization with derivatives}
The outer loop corresponds to a typical optimization problem with bound constraints and linear constrains as expressed in Sec. (\ref{outloop}). With the FDs derived in the Sec. (\ref{fds}), the constrained optimization problem can be implemented with the corresponding subroutines in the \textit{Scipy} library \cite{2020SciPy-NMeth}, for example, the "trust-region constrained method" or "sequential least squares programming algorithm".

The inner loop is a typical root search problem which can be solved numerically in various methods. The easiest one to implement is linear mixing method in which the $\boldsymbol{\mathcal{R}}$ matrix will be updated through mixing the input value and new output value at each step. Then it will be iterated many times from a given initial value until it's converged. However, it converges slowly and even fails to converge for some complex systems. Thus it is essential for us to derive a powerful newton method suitable for general scenarios. Define a new function as, 
\begin{equation}
\mathscr{F}(\boldsymbol{\mathcal{R}}) = \mathcal{I}(\boldsymbol{\mathcal{R}}) - \boldsymbol{\mathcal{R}}
\end{equation}
The solution of the inner loop is the root of $\mathscr{F}(\boldsymbol{\mathcal{R}})$, which can be solved by the standard Newton's method. First order of Taylor expansion near the root "$\boldsymbol{\mathcal{R}}_n$" gives us
\begin{equation}
\mathscr{F}(\boldsymbol{\mathcal{R}}_{n}) = \mathscr{F}(\boldsymbol{\mathcal{R}}_{n}) + \mathscr{F}'(\boldsymbol{\mathcal{R}}_n) \delta \boldsymbol{\mathcal{R}}_n = 0
\end{equation}
Thus
\begin{equation}\label{iln}
\boldsymbol{\mathcal{R}}_{n+1} = \boldsymbol{\mathcal{R}}_n +  \delta \boldsymbol{\mathcal{R}}_n = \boldsymbol{\mathcal{R}}_n - \mathscr{F}'(\boldsymbol{\mathcal{R}}_n)^{-1}  \mathscr{F}(\boldsymbol{\mathcal{R}}_n) \\
\end{equation}
with 
\begin{equation}
 \mathscr{F}'(\boldsymbol{\mathcal{R}}_n) = (\frac{\partial \boldsymbol{\mathcal{R}}_{n+1}}{\partial \boldsymbol{\mathcal{R}}}-1)\Big\vert_{\boldsymbol{\mathcal{R}}=\boldsymbol{\mathcal{R}}_n}
\end{equation}
where the subscript $n$ means the step when the iteration converges and the derivation of $\frac{\partial \boldsymbol{\mathcal{R}}_{n+1}}{\partial \boldsymbol{\mathcal{R}}}$ is shown in Eq. (\ref{prpr}). Therefore, we calculate $\boldsymbol{\mathcal{R}}$ as well as $\frac{\partial \boldsymbol{\mathcal{R}}_{n+1}}{\partial \boldsymbol{\mathcal{R}}}$ after each inner iteration and then $\boldsymbol{\mathcal{R}}$ will be updated through Eq. (\ref{iln}).

The search of the Lagrange multipliers including $\boldsymbol{\lambda^F}$ and $\boldsymbol{\lambda^B}$ can also be speeded up by the newton method. Take $\boldsymbol{\lambda^F}$ for example and the same for $\boldsymbol{\lambda^B}$. A new function of $\boldsymbol{\lambda^F}$ is defined as,
\begin{equation}
w(\boldsymbol{\lambda^F}) = \boldsymbol{n^0} - \boldsymbol{n^F}
\end{equation}

The search of $\boldsymbol{\lambda^F}$ which is tuned to satisfy the Eq. (\ref{cst1_n0}) equals to find the root of $w(\boldsymbol{\lambda^F})$. Following the derivation of the newton method of inner loop, it is easy to develop, 
\begin{equation}
\boldsymbol{\lambda}^F_{n+1} = \boldsymbol{\lambda}^F_{n} - (\frac{\delta w_n}{\delta \boldsymbol{\lambda}^F_n})^{-1}w(\boldsymbol{\lambda}^F_n) =  \boldsymbol{\lambda}^F_{n} +(\frac{\delta \boldsymbol{n}^F_n}{\delta \boldsymbol{\lambda^F}}_n)^{-1}w(\boldsymbol{\lambda^F}_n)
\end{equation}
where $\frac{\delta \boldsymbol{n}^F_n}{\delta \boldsymbol{\lambda}^F_n} = \frac{\partial_0  \boldsymbol{n}^F_n}{\partial_0 \boldsymbol{\lambda}^F_n}$ is the PPD of the Fermi part which will be derived in appendix. Besides, both $w$ and $\boldsymbol{\lambda^F}$ are real vectors so that the root search problem can also be done using the subroutines in the \textit{Scipy} library \cite{2020SciPy-NMeth}. Both the methods perform well and the latter one has the advantage in speed because root search subroutines in the \textit{Scipy} library do not need derivatives every iteration step.

\subsection{Extension to DFT+Gutzwiller}
DFT+G aims at solving the strongly interacting materials which are always described by a generalized tight-binding model including both correlated orbitals (COs) and non-correlated orbitals (NCOs). The Coulomb interaction among the electrons on the COs will be treated by the Gutzwiller variational method. When involving NCOs, there exists another constraint for the conservation of total electrons,
\begin{equation}
\sum_\beta n^0_\beta =  N_{total} - N_{corr} 
\end{equation} 
where $\beta$ labels the NCOs and $N_{total}$ is the number of total electrons.

That leads to the modified Fermi Hamiltonian,
\begin{equation}\label{mfh}
\begin{split}
\hat{H}^F =&  \sum_{i\ne j;\alpha,\beta,\delta,\gamma} t^{\alpha\beta}_{ij}\mathcal{R}^\dag_{\gamma\alpha} \hat{c}^\dag_{i\gamma} \hat{c}_{j\delta} \mathcal{R}_{\delta\beta}  + \sum_{i,\alpha=1} \lambda^{F,corr}_{\alpha} \ket{\alpha}\bra{\alpha} \\
&+ \sum_{i\ne j;\tilde{\alpha},\tilde{\beta}} t^{\tilde{\alpha}\tilde{\beta}}_{ij} \hat{c}^\dag_{i\tilde{\alpha}} \hat{c}_{j\tilde{\beta}}  +\sum_i \lambda^{F,uncorr} \sum_{\tilde{\beta}} \ket{\tilde{\beta}}\bra{\tilde{\beta}}
\end{split}
\end{equation}
where the $\alpha$ $(\gamma,\delta,\beta)$ in the first and second term traverses all the COs , the $\tilde{\alpha}$ $(\tilde{\beta})$ in the third and last part traverse all the NCOs. Here the original Lagrange multiplies of the Fermi part have beed redefined as $\boldsymbol{\lambda}^{F,corr}$ while $\lambda^{F,uncorr}$ denotes the Lagrange multiplies for NCOs. Thus the Lagrange multiplies of Fermi part are unified as $\boldsymbol{\lambda^F} = (\boldsymbol{\lambda}^{F,corr},\lambda^{F,uncorr})$.

Obviously, the optimization of total energy in the outer loop is just with respect to the variational single particle density matrix of COs because the total number of electrons on the NCOs is fixed throughout the whole process.

Besides, some real materials are not able to be characterized by a tight-binding model, especially for those containing f orbitals. Fortunately, the information on COs can be encoded by the so-called $S$-matrix \cite{deng2009local}, which is the overlap matrix between Bloch states and the local orbitals at each k point. The definition reads,
\begin{equation}
S_{\boldsymbol{k},\alpha n} = \bra{\boldsymbol{k}\alpha}\ket{\boldsymbol{k}n} , 
\end{equation} 
with 
\begin{equation}
\ket{\boldsymbol{k}\alpha} = {1 \over \sqrt{N_R}} \sum_j e^{i\boldsymbol{k}\cdot\boldsymbol{R}_j} \ket{j,\alpha}
\end{equation}
where $\ket{\boldsymbol{k}n}$ is the eigenfunction of the $n$-th band with momentum $\boldsymbol{k}$. $N_R$ is the number of lattice sites in real space. All the quantities involved in the Fermi part are formulated easily in $\boldsymbol{k}$-space with the help of $\boldsymbol{S_k}$ matrix. (See the Appendix for details.)

When combining DFT with Gutzwiller method or DMFT, the so-called double counting (DC) energy should always be removed because the correlation effect has already been partially considered in the exchange correlation functional like local density approximation (LDA) and generalized gradient approximation (GGA) functional. Thus a term $\hat{H}_{dc}$ should be subtracted from the original Hamiltonian. There are different numerical schemes to remove the DC energy. In most of these schemes, the DC correction only changes the total occupation numbers for the correlated orbitals ($N_{total}$). Therefore the DC correction term can be absorbed into $\boldsymbol{\lambda^F}$ which is determined by enforcing the calculated occupation number to match the given $\boldsymbol{n^0}$.

\section{Atomic problem}
\label{gvp}
In this section, we will introduce the atomic problem, which is essential for the construction of the Gutzwiller projector. Based on the symmetry of atomic eigenstates, we can design the Gutzwiller variational parameters.

The local atomic hamiltonian takes the form,
\begin{equation}
\hat{H}_i^{at} = \hat{H}_i^{int} + \hat{H}_i^{cf} + \hat{H}_i^{soc}
\end{equation} 
where $\hat{H}^{int}$ denotes the onsite electron-electron Coulomb interaction, $\hat{H}^{cf}$ is the crystal field splitting (CF) and $\hat{H}^{soc}$ is the spin-orbital coupling (SOC) term. For rare earth elements such as lanthanides, both the CF and SOC play an important role. As for the Coulomb interaction, we adopt the rotationally invariant Kanamori form, which reads

\begin{equation}\label{kanamori}
\begin{split}
\hat{H}^{int}_i =& \frac{U}{2}\sum_{a,\sigma\ne\sigma'} \hat{n}_{ia\sigma} \hat{n}_{ia\sigma'}  + {U' \over 2} \sum_{a\ne b,\sigma\ne\sigma'}\hat{n}_{ia\sigma} \hat{n}_{ib\sigma'} \\
 &+ (U'-J)\sum_{a>b,\sigma} \hat{n}_{ia\sigma}\hat{n}_{ib\sigma} 
-J\sum_{a\ne b,\sigma\ne\sigma'} \hat{c}^\dag_{ia\sigma}\hat{c}^\dag_{ib\sigma'}\hat{c}_{ib\sigma}\hat{c}_{ia\sigma'}\\
& - J\sum_{a\ne b,\sigma\ne\sigma'} \hat{c}^\dag_{ia\sigma}\hat{c}^\dag_{ia\sigma'}\hat{c}_{ib\sigma}\hat{c}_{ib\sigma'}
\end{split}
\end{equation}

where $\hat{c}^\dag_{ia\sigma} (\hat{c}_{ib\sigma'})$ represents the creation (annihilation) of an electron on a (b) orbital with $\sigma$ $(\sigma')$ spin component of site $i$. The first three terms denote the density-density Coulomb interaction. The last two describe the spin-flipping and pari-hopping interaction which are important for unfilled orbitals with more than one electron. The site index $i$ will be dropped off hereafter.

After diagonalizing the atomic hamiltonian in Fock space, we get the atomic eigenstates ($\ket{\Gamma}$) and eigenvalues ($E_{\Gamma}$),
\begin{equation}
\hat{H}^{at} \ket{\Gamma} = E_{\Gamma} \ket{\Gamma}
\end{equation}

The Gutzwiller trail wavefunction is defined by the Gutzwiller projector ($\hat{P}$) acting on the non-interacting wave function. Hence it's necessary for us to figure out the symmetry of each atomic eigenstate in order to construct proper $\hat{P}$. For continuous symmetries, all the corresponding conserved quantities consist of a complete set of commuting observables (CSCO). For example, the electron system without SOC and CF has U(1) $\otimes$ SO(3) $\otimes$ SU(2) symmetry, which leads to the CSCO of $\{\hat{H},\hat{N},\hat{L}^2,\hat{L}_z,\hat{S}^2,\hat{S}_z\}$. If the Hamiltonian is independent of time (the current case), the eigenstates can be labelled by the good quantum numbers (GQNs) $\{N,L,L_z,S,S_z\}$. When it comes to the discrete symmetries, like the double point group, the eigenstates can be labelled by its irreducible representations (IRs). This is the case for the rare earth compounds, where both the SOC and CF play important roles.

Given a system subject to the double point group $G$ and with no accidental degeneracy, we need to label each atomic eigenstate by IRs in the target Fock space, 
\begin{equation}
\boldsymbol{U} = \{\psi^{q,s}_m, \cdots \}
\end{equation}
where $q$ denotes different IRs of $G$, $s$ represents different $q$-th IRs when its multiplicity ($r$) is larger than one and $m$ labels the bases belonging to $q$-th IR. Then the atomic Hamiltonian matrix ($\boldsymbol{H}^{at}$) can be diagonalized by an unitary transformation,
\begin{equation}
\boldsymbol{H}^{at}_{diag} = \boldsymbol{U}^{-1} \boldsymbol{H}^{at} \boldsymbol{U}
\end{equation}
where $\boldsymbol{H}^{at}_{diag}$ is a diagonal matrix whose elements are eigenvalues of $\boldsymbol{H}^{at}$.

How to obtain $\psi^{q,s}_m$? The projection operator in the group theory (POG) \cite{altmann1994point} works. For sake of convenience, the matrix representation of $\hat{g} \in G$ in each target Fock space of different particle numbers shall be prepared in advance. Now we use the formula below to determine the multiplicities of IRs in reduction at first,
\begin{equation}
r^q = {1 \over |G|} \sum_{\hat{g} \in G} \chi^{q,*}(\hat{g}) \chi(\hat{g})
\end{equation} 
where $|G|$ is the rank of group $G$ which contains $\hat{g}$ as elements. $\chi^q(\hat{g})$ is the character of $\hat{g}$ for $q$-th IR while $\chi(\hat{g})$ is that of the original reducible matrix representation.

Then the POG is constructed as 
\begin{equation}
\hat{W}^{q}_{np} = \frac{|\check{G}^{q}|}{|G|} \sum_{\hat{g} \in G} \check{G}^{q,*}_{np}  \hat{g}  
\end{equation}
where $\check{G}^{q}$ is the matrix representation of $q$-th IR whose dimension is $|\check{G}^{q}|$. $\hat{W}^{q}_{np}$ has the property to project out the $n$-th basis of $q$-th IR at a given $p$ index,
\begin{equation}
\hat{W}^{q}_{np} \psi^{q'}_{p'} = \psi^q_n \delta_{qq'}\delta_{pp'} 
\end{equation}
Thus all the bases of $q$-th IR with nonvanishing multiplicity can be projected out by applying $\hat{W}^{q}_{21}$ on a random initial function containing $\psi^{q}_{1}$ and then applying $\hat{W}^{q}_{32}$ on the $\psi^{q}_{2}$ and so on until the index $n$ has ranged over all the dimension of $q$-th IR. When it comes to the IR with multiplicity larger than unit, it just needs to choose another proper initial function which results in the basis linearly independent from the previous bases followed by repeating the projection process again. However, the basis set of different multiplicity are not orthogonal to each other. An alternative way to cope with this is to just orthogonalize the first basis belonging to different multiplicities through Schmidt orthogonalization and then the unitary transformation applies to all the other columns as well.

Upon obtaining all the atomic eigenstates with definite symmetry $\Gamma_{z,q,s,m}$ ($z$ represents GQNs, $s$ is used to label different $q$-th IR whose multiplicity $r$ is larger than one and $m$ denotes bases belonging to different columns of $q$-th IR), the Gutzwiller projector can be designed to meet our specific demands. For example, if we want to investigate the ground state with original symmetry, the Gutzwiller projector (the site index $q$ has been dropped for simplicity) is defined as 
\begin{equation}
\hat{P} = \sum_{\Gamma\Gamma'} \lambda_{\Gamma\Gamma'} \ket{\Gamma_{z,q,s,m}}\bra{\Gamma'_{z',q',s',m'}} \delta_{zz'} \delta_{qq'} \delta_{mm'}
\end{equation}
which is named as "symmetric Gutzwiller projector (SGP)". Besides, the constraints by Kronecker delta function could be selectively loosened when we are interested in some symmetry breaking state. 

The other two frequently used Gutzwiller projectors are diagonal Gutzwiller projector (DGP) and another most general Gutzwiller projector (GGP). The former denotes the projector which only contains the diagonal matrix elements in the atomic basis,
\begin{equation}
\hat{P} = \sum_{\Gamma\Gamma'} \lambda_{\Gamma\Gamma'} \ket{\Gamma_{z,q,s,m}}\bra{\Gamma'_{z',q',s',m'}} \delta_{zz'}\delta_{qq'}\delta_{ss'}\delta_{mm'}
\end{equation}
The latter means the projector contains matrix elements between arbitrary atomic eigenstates with the same occupation number, which reads
\begin{equation}
\hat{P} = \sum_{\Gamma\Gamma'} \lambda_{\Gamma\Gamma'} \ket{\Gamma_{n,q,s,m}}\bra{\Gamma'_{n',q',s',m'}} \delta_{nn'}
\end{equation}
where $n$ $(n')$ denotes the particle number of configurations here.

\section{Implementation}\label{imple}

\begin{figure}[ht]
\includegraphics[width=0.49\textwidth]{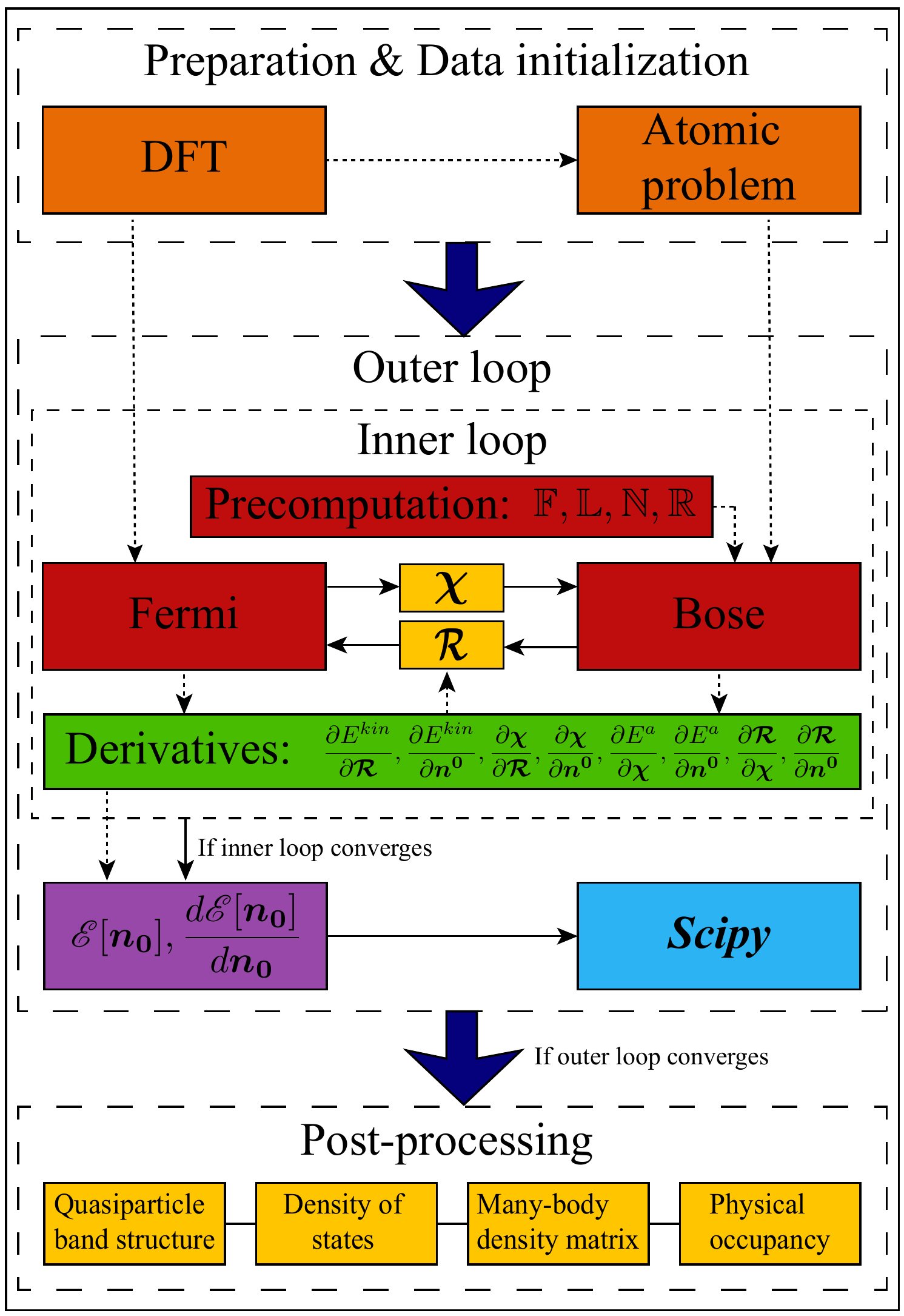}
\centering
\caption{\label{Fig:fb5}(Color online). The hierarchical structure of the implementation of DFT + G. There are three parts in total. The first one in orange contains DFT interface and atomic subroutine. The second one in the middle of the flow chart is the key part of DFT + G including both outer and inner loops. The final one is post-processing to obtain quasiparticle band structure, density of states and so on. The big blue arrows represents the workflow to perform the calculations. The dashed lines with arrow indicate the data transfer direction. And the solid line with arrow means the iteration process. }
\end{figure}

How to implement DFT + G will be present in this section. As illustrated in Fig. \ref{Fig:fb5}, the entire implementation is divided into three parts. For data initialization, the interface to DFT should be prepared to get the information the local orbitals from Bloch states. If the material can be fitted by a tight-binding model, the interface is just to obtain the Bloch states by diagonalizing the single particle Hamiltonian and then project the Bloch states onto the local orbitals. If not, the interface can be directly embedded into the \textit{ab initio} software such that the overlap matrix ($S$-matrix) is naturally computed. Finally, an atomic subroutine taking symmetry into consideration for Bose part needs to be programmed.

The second part containing outer loop and inner loop is of greatest importance throughout the entire implementation. As for inner loop, some many-body matrices, like $\mathbb{F}, \mathbb{L}, \mathbb{N}$ and $\mathbb{R}$, remain unchanged for given $\boldsymbol{n^0}$ during the iterative process. Thus, it's efficient to do one-shot calculation for them before the inner iteration starts. After receiving the information of local orbitals and energy bands from DFT calculations as well as the input parameter $\boldsymbol{\mathcal{R}}$, the first equation of Eq. (\ref{se2}) can then be solved and the single particle energy as well as the corresponding PDs, namely $\frac{\partial E^{kin}}{\partial \boldsymbol{\mathcal{R}}},\frac{\partial E^{kin}}{\partial \boldsymbol{n^0}},\frac{\partial \boldsymbol{\chi}}{\partial \boldsymbol{\mathcal{R}}},\frac{\partial \boldsymbol{\chi}}{\partial \boldsymbol{n^0}}$, in the greed box of Fig. \ref{Fig:fb5} can be calculated. Similarly, the many-body effective Hamiltonian is able to be constructed with the data from atomic subroutine and pre-calculated many-body matrices as well as the intermediate parameter $\boldsymbol{\chi}$. Thus the atomic energy and the corresponding PDs such as $\frac{\partial E^a}{\partial \boldsymbol{\chi}},\frac{\partial E^a}{\partial \boldsymbol{n^0}},
\frac{\partial \boldsymbol{\mathcal{R}}_{n+1}}{\partial \boldsymbol{\chi}},\frac{\partial \boldsymbol{\mathcal{R}}_{n+1}}{\partial \boldsymbol{n^0}}$ can be computed after solving the second one of Eq. (\ref{se2}). Fermi part and Bose part together form an iterative loop sketched by the red box connected through the solid lines with arrows.
After inner loop converges, the total energy and TD are naturally obtained according to Eq. (\ref{op1}) and Eq. (\ref{td}), which will be passed to the subroutines in the \textit{Scipy} Library to perform constrained minimization.

The last part is the post processing, where the quasi-particle band structure, density of states, many body density matrix and the orbital occupancy can be easily computed using the converged GWF under the GA.

\section{Benchmark Results}
\label{bcmk}

In this section, two correlated models will be calculated as benchmarks for our method. The first model is a two-band Hubbard model with diagonal crystal field and the other one is the doped bilayer Hubbard model. Besides, three types of Gutzwiiler projector will be compared for two different point group symmetries.

\subsection{Two-band Hubbard model}

The tight-binding model reads,
\begin{equation}
t_{\boldsymbol{k}}^{ab} = -{1 \over 3} t^0_{ab} \sum_{u=1}^3 \cos k_\mu
\end{equation}
where $t^0_{ab} = \delta_{ab}$. The crystal field takes the simple form, 
\begin{equation}
\hat{H}_{cf} = \sum_{\sigma} \Delta(\hat{n}_{1\sigma} - \hat{n}_{2\sigma})
\end{equation}
with $\Delta = 0.2$. The orbital index "1" represents anti-bonding state while "2" denotes bonding state. The Coulomb interaction employs the Kanamori form introduced in Eq. (\ref{kanamori}) with $J/U=0, 0.01, 0.02, 0.05, 0.10, 0.15, 0.25$, respectively. This model has been explored by the Gutzwiller method proposed in Ref. \cite{lanata2012efficient} as well as DMFT \cite{werner2007high}, which are used for benchmark here.

\begin{figure}[ht]
\includegraphics[width=0.45\textwidth]{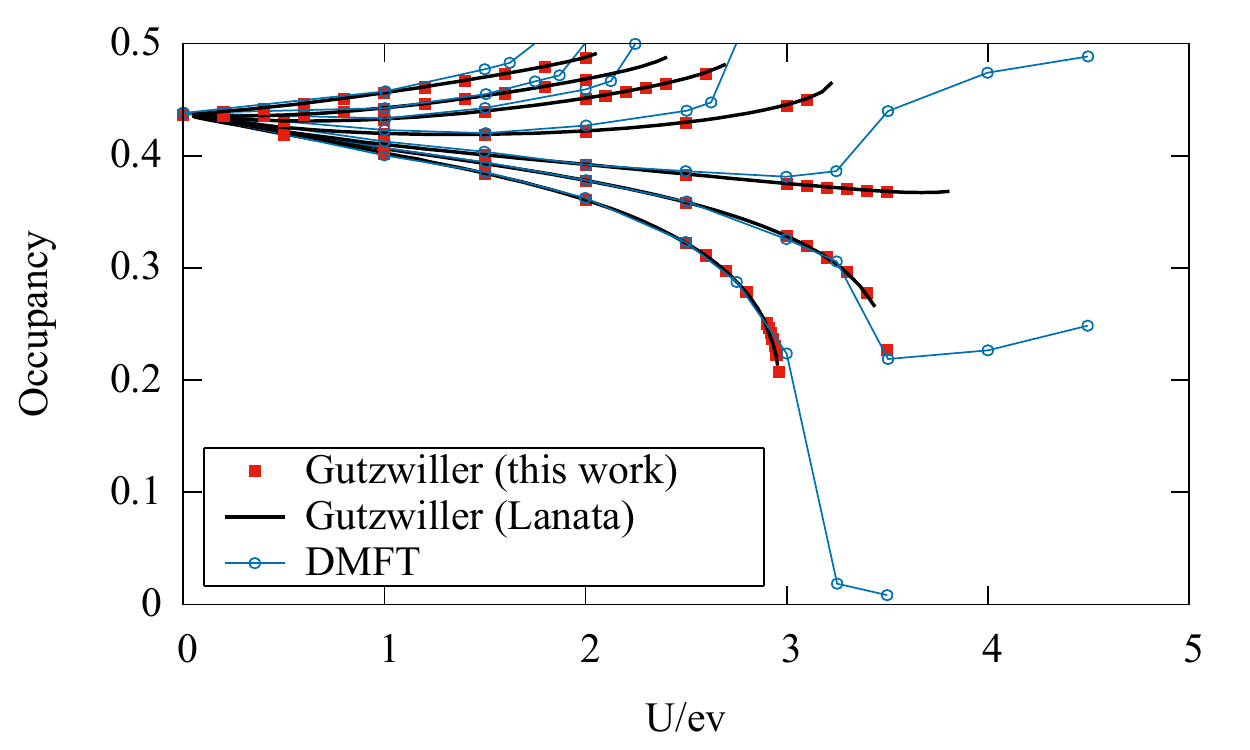}
\centering
\caption{\label{tbho}(Color online). Occupancy of orbital 1 for crystal field splitting $\Delta=0.2$ and different $J/U$ values (0, 0.01, 0.02, 0.05, 0.10, 0.15, 0.25 from bottom to top). The red square dots are calculated by the Gutzwiller method in the present paper and the continuous black lines refer to the Gutzwiller algorithm in Ref. \cite{lanata2012efficient} and the blue linepoints data refer to the results by DMFT \cite{werner2007high}.}
\end{figure}

In order to compare with other numerical methods, a semicircular density of states is adopted in the Fermi part. For Bose part, there is no cut off of the Fock space and the GGP (70 parameters in total) is adopted.

\begin{table*}
\centering
\caption{\label{tbhtb}The information of convergence as well as the comparison between the analytical derivatives and numerical derivatives for two-band Hubbard model with $\Delta=0.2$ at $U=2.5$ and various J/U values. AD and ND are calculated at the step when it converges. Some abbreviations used for brevity are explained below. "AD" ("ND") means analytical total derivatives (numerical total derivatives). “TE” refers to the optimization of total energy in the outer loop while "$\mathcal{R}$" refers to the fixed point problem of $\mathcal{R}$ in the inner loop. "AD-Newton" denotes the newton method with analytical derivatives while "LM" denotes the linear mixing update approach.}\label{tab1}
\begin{tabular}{ccccccc}
\toprule
\multirow{2}*{$J/U$}&\multirow{2}*{AD} & \multirow{2}*{ND} & \multicolumn{4}{c}{Number of Steps} \\
\cmidrule{4-7}
&&& TE (AD) & TE (ND)    & $\mathcal{R}$ (AD-Newton) & $\mathcal{R}$ (LM) \\
\midrule
0&6.000036&6.000379&5&70&15&15 \\
0.01&5.900268&5.900586&5&23&14&15 \\
0.02&5.800821&5.801132&5&9&13&16 \\
0.05&5.499973&5.500264&6&21&13&14 \\
0.10&4.999945&5.000238&6&30&15&18 \\
0.15&4.501501&4.501818&5&22&18&26 \\
0.25&3.528302&3.528777&2&6&37&69 \\
\bottomrule
\end{tabular}
\end{table*}

The occupancy in Fig. \ref{tbho} is not the variational density matrix ($\boldsymbol{n^0}$) but the physical occupancy calculated through Eq. (\ref{ob3}). The results by the present Gutzwiller method matches exactly with that by another implementation of the Gutzwiller method \cite{lanata2012efficient}. Both of the two Gutzwiller results are consistent well with the DMFT results in the metallic phases (small U region).

The comparison between the numerical total derivatives (ND) and analytical total derivatives (AD) of the orbital 1 for different $J/U$ value is listed in  Table \ref{tbhtb}. The ND is obtained by the finite difference approach with a step size of $n^0_1 * 10^{-4}$. All the AD and the ND match very well with each other by a relative error of about $10^{-5}$, which proves the validity of the analytical formulas of all the process derivatives. The efficiency of different numerical methods for optimization and root searching is also shown. The precision goal of the minimization in the stopping criterion is set to be $10^{-10}$ while that of the root searching in the inner loop is set to be $10^{-12}$. As for the minimization of the total energy in outer loop, the subroutine in \textit{Scipy} with AD and ND has different performance. The optimization with AD has very stable performance with the number of steps just about 5. However, the optimization with ND achieves at a maximum number of 70 because it always fluctuates unpredictably which makes the optimization expensive or even failed. The finding of root of $\mathscr{F}(\boldsymbol{\mathcal{R}})$ in the inner loop is implemented by the newton method and linear mixing approach respectively. From the last two columns of the table, we know that the two approaches have similar rate of convergence at J/U  equals from 0 to 0.1. But the newton method with AD converges at a faster speed than the linear mixing approach when J/U is greater than 0.1. Therefore, the AD proves to be very powerful in the minimization and root search process.

\subsection{Bilayer Hubbard model}

The bilayer Hubbard model can be viewed as a special type of the two-band Hubbard model described in the previous section with the following form of the crystal field, 
\begin{equation}
\Delta = 
\begin{pmatrix}
0&0&V&0\\
0&0&0&V\\
V&0&0&0\\
0&V&0&0
\end{pmatrix}
\end{equation}
where V=0.25 is the hybridization term between the two layers. As for the Coulomb interaction, $U' = J = 0$ is assumed and the parameter $U$ is scanned from 0 to 3.5. The onsite hybridized term $\Delta$ between the bilayers makes the quasi-particle weight matrix ($Z$) and the density matrix non diagonal. With the unitary transformation

\begin{figure}[h]
\includegraphics[width=0.49\textwidth]{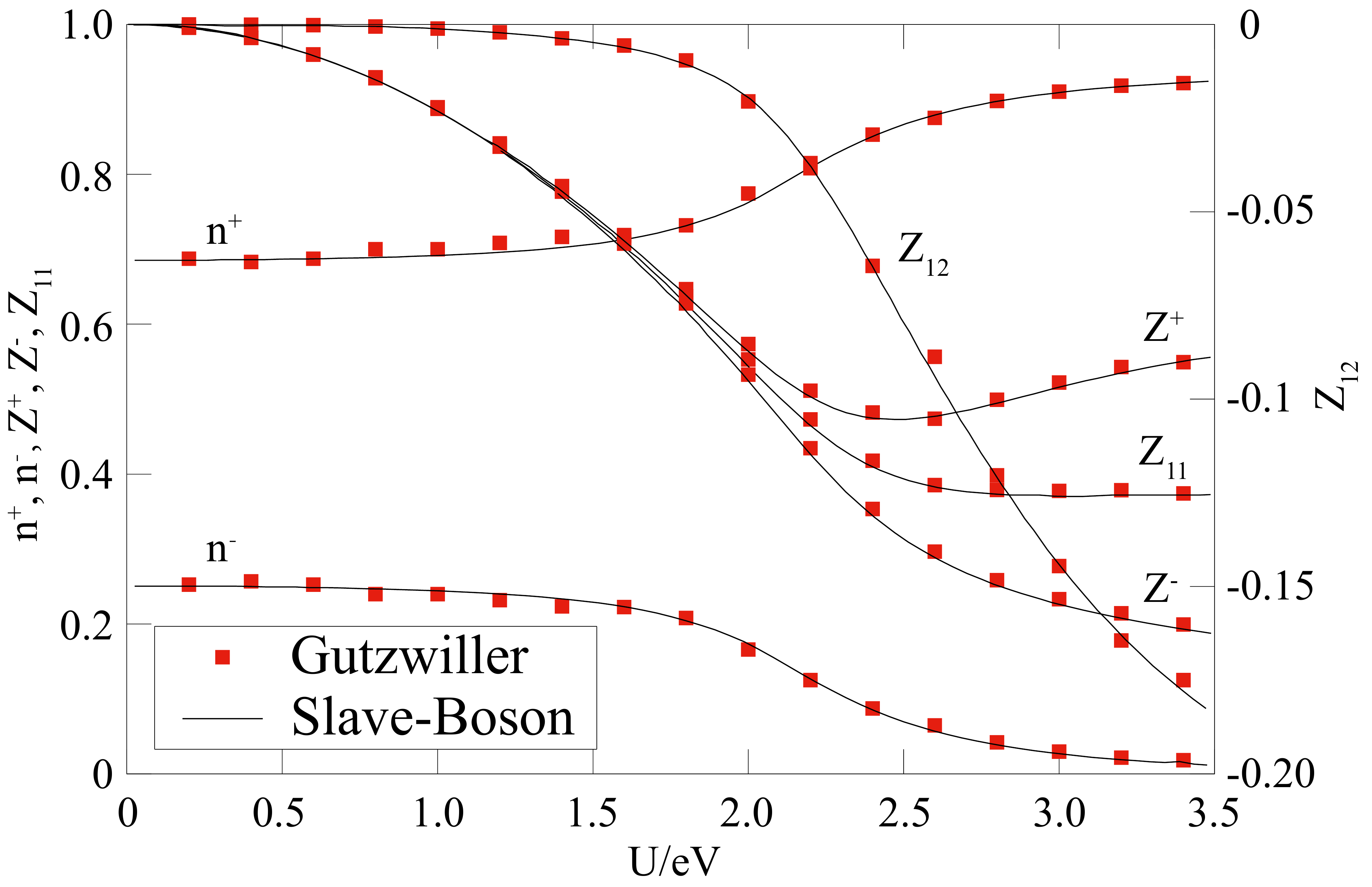}
\centering
\caption{\label{bho}(Color online). The quasi particle weight and fillings for the doped bilayer Hubbard model ($n=1.88$) with equal bandwidths and the crystal field splitting $V=0.25$.}
\end{figure}

\begin{equation}
\boldsymbol{U}={1 \over \sqrt{2}} 
\begin{pmatrix}
1 &0&-1&0\\
0&-1&0&1\\
-1&0&-1&0\\
0&1&0&1
\end{pmatrix},
\end{equation}
the natural bases have been well defined under which $Z$ and density matrix take the diagonal form
\begin{equation}
\boldsymbol{Z} =
\begin{pmatrix}
Z^+ & 0 & 0& 0\\
0 & Z^+ & 0& 0\\
0 & 0& Z^- &0\\
0&0&0&Z^-  \\
\end{pmatrix}
=
\boldsymbol{U}^{-1} \begin{pmatrix}
Z_{11}&0&Z_{12}&0\\
0&Z_{11}&0&Z_{12}\\
Z_{21}&0&Z_{22}&0\\
0&Z_{21}&0&Z_{22}
\end{pmatrix}\boldsymbol{U}
\end{equation}
with $Z_{11} = Z_{22}$ and $Z_{12} = Z_{21}$, and 
\begin{equation}
\boldsymbol{n^0} = 
\begin{pmatrix}
n^{+} &0&0&0\\
0&n^{+} &0&0\\
0&0&n^{-}&0\\
0&0&0&n^{-}\\
\end{pmatrix}
\end{equation}
where the superscript $"+/-"$ denotes the bonding/anti-bonding orbitals.

\begin{figure}[th]
\includegraphics[width=0.45\textwidth]{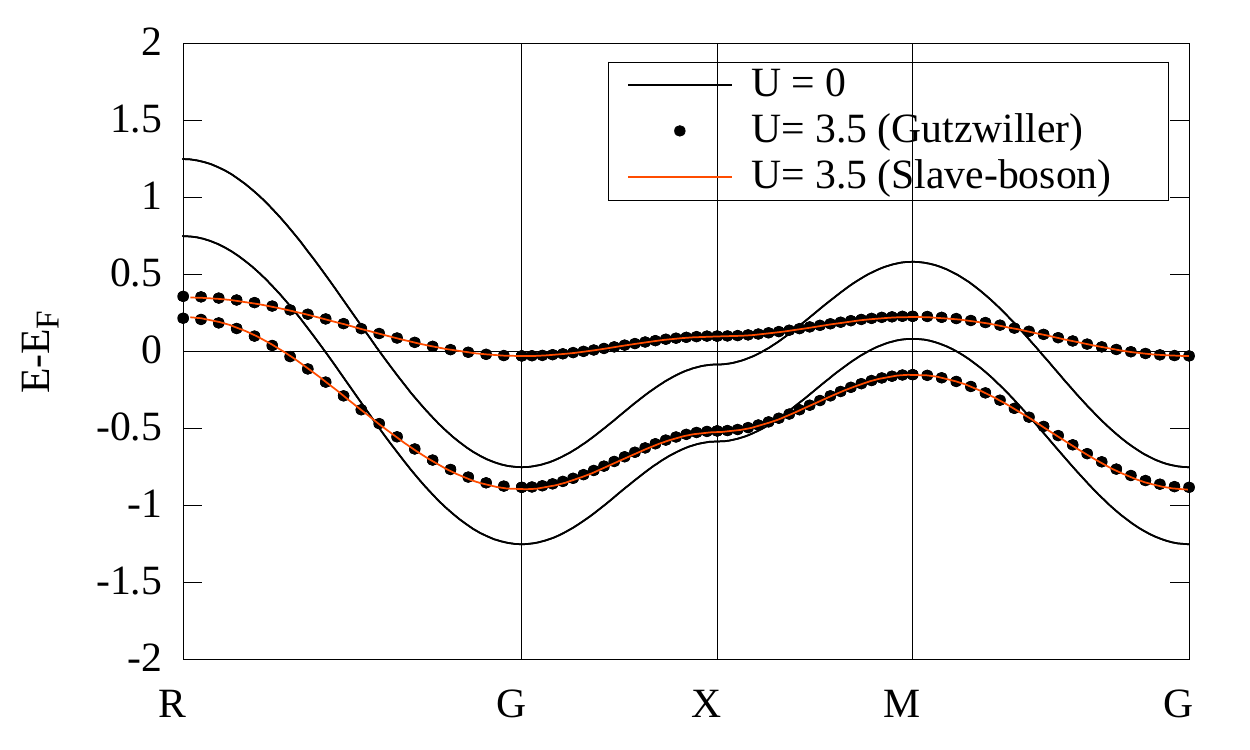}
\centering
\caption{\label{bhb}(Color online). Quasi particle bands of the doped bilayer Hubbard model ($n=1.88$) with equal bandwidths and crystal field splitting $V=0.25$.}
\end{figure}
The total occupancy per site is assumed to be 1.88 following Ref. \cite{lechermann2007rotationally} and the noninteracting DOS of cubic lattice is adopted. For Bose part, we also include the complete Fock space and all the 70 GGP parameters are kept.
\begin{table*}
\label{atomic}
\centering
\caption{The reduction for representations of the double point group (DPG) D$_{4h}$ and D$_{2h}$ in Fock space of $N=1\sim3$ of f-orbital systems and the corresponding three different types of the Gutzwiller projector. The notation of IRs from $GM_1^+$ to $\bar{GM_9}$ in the second row is according to Ref. \cite{cracknell1979general}. $GM_1^+$ to $\bar{GM_6}$ are one dimensional IR while $\bar{GM_8}$ and $\bar{GM_9}$ are two dimensional IR. The second column of the table means Fock spaces of different particle numbers. The numbers in the third to tenth column denote the multiplicity of IR. The data in the last three columns are the number of parameters in DGP, GGP and SGP respectively.}\label{tab2}
\begin{tabular}{ccccccccccccc}
\toprule
\multirow{2}*{\text{DPG}}&\multirow{2}*{$N$} & \multicolumn{8}{c}{Irreducible Represenation} &\multicolumn{3}{c}{Gutzwiller Projector} \\
\cmidrule{3-10}  \cmidrule{11-13}
&&$GM_1^+$&$GM_2^+$&$GM_3^+$&$GM_4^+$&$GM_5^+$&$\bar{GM_6}$&$\bar{GM_8}$&$\bar{GM_9}$&DGP&GGP&SGP \\
\midrule
\multirow{3}*{D$_{4h}$}&1&0&0&0&0&0&0&4&3&14&196&50\\
&2&16&12&9&12&21&0&0&0&91&8281&1507\\
&3&0&0&0&0&0&0&91&91&364&132496&33124\\
\cmidrule{2-13}
\multirow{3}*{D$_{2h}$}&1&0&0&0&0&0&7&0&0&14&196&98\\
&2&28&21&21&21&0&0&0&0&91&8281&2107\\
&3&0&0&0&0&0&182&0&0&364&132496&66248\\
\bottomrule
\end{tabular}
\end{table*}
The physical quantities of the ground states for U ranging from 0 to 3.5 are shown in Fig. \ref{bho}. The filling of anti-bonding states decreases monotonically with U and so do $Z^{-}, Z_{11}$ and $Z_{12}$. However, $Z^{+}$ decreases steadily to about 0.47 at first stage which is then followed by a slight increase at $U\approx2.4$. Besides, the non-interacting bands ($U = 0 $) and the quasi particle bands for $U=3.5$ calculated by Gutzwiller method as well as slave-boson method are illustrated in Fig. \ref{bhb}. The results of the proposed Gutzwiller method match perfectly with that calculated by the slave-boson mean field method.

\subsection{Examples for different Gutzwiller projectors}
Sec. \ref{gvp} provides a route to construct the Gutzwiller projector subject to a certain point group symmetry. Although SGP is not used in the two models above due to their simplicity, it's essential for us to employ SGP for some real materials with d- or f-orbital. Here, the comparison between SGP and the other two types of Gutzwiller projectors for two different point groups in f shell are present in this section. Both the crystal field and the spin-orbit coupling are very important in these systems, so the double point group is considered here.

Take the systems with f-orbital subject to $D_{4h}$ and $D_{2h}$ double point group for example. The details of reduction of the target Fock space ($N=1\sim3$) are listed in the middle eight columns of Table \ref{atomic}. With these information, the numbers of parameters in the three types of Gutzwiller projector are easily figured out, which are also listed in the last three columns. From the table, it's obvious that the number of parameters in SGP is much less than that in GGP after taking the double point group symmetry into consideration. Although DGP has less Gutzwiller parameters than DGP, it is at the price of losing accuracy. Therefore, compared with GGP and DGP, SGP shows a great balance between the efficiency and accuracy.

\section{Conclusion}
\label{ccls}
In this work, we have developed a very efficient solver to solve the Gutzwiller variational problem with general form of the interaction. The ground state at fixed single particle variational density matrix ($\boldsymbol{n^0}$) will be firstly obtained in the inner loop which is then followed by an optimization of the total energy with respect to $\boldsymbol{n^0}$ in the outer loop until the total energy reaches a minimum. As a starting point, a tight-binding model or the overlap matrix (the $\boldsymbol{S}$-matrix) between Bloch states and the local orbitals are required as the input.

There are two key points in this paper. Firstly, all the analytical derivatives of the implementation processes of this algorithm have been derived through perturbation theory and the chain rule. These derivatives improve the efficiency of the convergence of both the outer loop and inner loop, sometimes even enable the cases which fail to converge through the old scheme convergent. In addition, the searching for Lagrange multipliers of both Fermi and Bose part can also benefit from the analytical derivatives. Secondly, an atomic algorithm with the full point group consideration has been set up for the design of the Gutzwiller projector. Each atomic eigenstate is thus labelled by the GQNs and IRs, which enables us to customize Gutzwiller projector subject to a desired symmetry.

The two-band Hubbard model with diagonal crystal field and the bilayer Hubbard model have been studied by the present method as the benchmarks. Both cases are in good agreement with the previous published results by DMFT, multi-orbital slave-boson mean field method or the previously developed Gutzwiller method. Three different types of the Gutzwiller projector, namely GGP, DGP and SGP, are compared for f-orbital with $D_{4h}$ and $D_{2h}$ double point group symmetry. It shows that SGP contains significantly less parameters than GGP without losing any symmetry allowed component, which leads to a perfect balance between the accuracy and efficiency.

\section{Acknowledgments}
Hongming Weng acknowledges the supports from the National Natural Science Foundation (Grant No. 11925408), the Ministry of Science and Technology of China (2018YFA0305700), the Chinese Academy of Sciences (Grant No. XDB33000000), the K. C. Wong Education Foundation (GJTD-2018-01), the Beijing Natural Science Foundation (Z180008), and the Beijing Municipal Science and Technology Commission (Z191100007219013). Shiyu Peng thanks Liang Du and Chuang Chen for helpful discussions.

\end{small}

 \appendix

 \section{Hamiltonian of the Fermi part in $\boldsymbol{k}$-space}
 \label{appendixA}
 A general tight-binding model contains both COs and NCOs which valid the completeness relation \cite{deng2009local},
 \begin{equation}
 \sum_{i,\alpha} \ket{i\alpha}\bra{i\alpha} + \sum_{i,\tilde{\alpha}} \ket{i\tilde{\alpha}}\bra{i\tilde{\alpha}} = 1
 \end{equation}

Taking the Coulomb interaction and Gutzwiller approximation into consideration, the COs will be renormalized such that the Hamiltonian turns to be 
\begin{widetext}
\begin{equation}
\begin{split}
\hat{H}^F =& \Big( \sum_{i,\alpha} \ket{i\alpha}\mathcal{R}^\dag_{i,,\alpha\beta}\bra{i\beta} + \sum_{i,\tilde{\alpha}} \ket{i\tilde{\alpha}}\bra{i\tilde{\alpha}}\Big)\hat{H}^{LDA}\Big( \sum_{i,\alpha} \ket{i\alpha}\mathcal{R}_{i,\alpha\beta}\bra{i\beta} + \sum_{i,\tilde{\alpha}} \ket{i\tilde{\alpha}}\bra{\tilde{i\alpha}}\Big) + \sum_{i,\alpha} \lambda^{F,corr}_{\alpha} \ket{\alpha}\bra{\alpha}  \\
&+ \sum_{i} \lambda^{F,uncorr} \sum_{\tilde{\alpha}} \ket{\tilde{\alpha}}\bra{\tilde{\alpha}}  \\
=& \Big( \sum_{i,\alpha} \ket{i\alpha}\mathcal{R}^\dag_{i,\alpha\beta}\bra{i\beta} + 1 - \sum_{i,\alpha} \ket{i\alpha}\bra{i\alpha}\Big)\hat{H}^{LDA}\Big( \sum_{i,\alpha} \ket{i\alpha}\mathcal{R}_{i,\alpha\beta}\bra{i\beta} + 1-  \sum_{i,\alpha} \ket{i\alpha}\bra{i\alpha}\Big) + \sum_{i,\alpha} \lambda^{F,corr}_{\alpha} \ket{\alpha}\bra{\alpha} \\ 
&+ \sum_{i} \lambda^{F,uncorr} (1-\sum_{{\alpha}} \ket{{\alpha}}\bra{{\alpha}})
\end{split}
\end{equation}
\end{widetext}
where the onsite part is absent from $\hat{H}^{LDA}$. The Hamiltonian in momentum space is easily derived by Fourier transformation $\ket{\boldsymbol{k}\alpha} = {1 \over N} \sum_i e^{i\boldsymbol{k}\cdot\boldsymbol{R_i}}\ket{i\alpha}$, 
\begin{equation}
\begin{split}
\hat{H}^F_{\boldsymbol{k}} =& (1- \hat{P}_{\boldsymbol{k}} + \hat{Q}_{\boldsymbol{k}}) \hat{H}^{LDA}_{\boldsymbol{k}} (1- \hat{P}_{\boldsymbol{k}} + \hat{Q}_{\boldsymbol{k}})\\
&+ \sum_{\alpha} \lambda^{F,corr}_{\alpha} \ket{\boldsymbol{k}\alpha}\bra{\boldsymbol{k}\alpha} +\lambda^{F,uncorr} (1- \hat{P}_{\boldsymbol{k}})
\end{split}
\end{equation}
with 
\begin{equation}
\begin{split}
\hat{P}_{\boldsymbol{k}} &= \sum_\alpha \ket{\boldsymbol{k}\alpha}\bra{\boldsymbol{k}\alpha} \\
\hat{Q}_{\boldsymbol{k}} &= \sum_{\alpha\beta} \ket{\boldsymbol{k}\alpha}\mathcal{R}_{\alpha\beta}\bra{\boldsymbol{k}\beta} \\
\end{split}
\end{equation}
When adopting the original Bloch representation, $\hat{P}_k$ and $\hat{Q}_k$ is encoded by $S$ matrix, 
\begin{equation}
\begin{split}
P_{k,mn}  &= \sum_\alpha S_{k,\alpha m}^\dag S_{k,\alpha n} \\
\boldsymbol{Q}_{k} &=  \boldsymbol{S}^\dag_{k} \boldsymbol{\mathcal{R}} \boldsymbol{S}_k
\end{split}
\end{equation}

 \section{Derivation of orbital renormalization matrix $\boldsymbol{\mathcal{R}}$}
The matrix $\boldsymbol{\mathcal{R}}$ comes from the renormalization of the many-body effects on the hopping term such that we will start from the kinetic energy,
\begin{equation}
\begin{split}
E^{kin} &= \sum_{i\ne j,\alpha,\beta} t^{\alpha\beta}_{ij} \bra{G}\hat{c}^\dag_{i\alpha}\hat{c}_{j\beta}\ket{G}\\
&= \sum_{i\ne j,\alpha,\beta} t^{\alpha\beta}_{ij} \bra{0}\hat{P}^\dag \hat{c}^\dag_{i\alpha}\hat{c}_{j\beta}\hat{P}\ket{0}\\
&= \sum_{i\ne j,\alpha,\beta} t^{\alpha\beta}_{ij} \bra{0}\hat{P}^\dag_i \hat{c}^\dag_{i\alpha}\hat{P}_i \hat{P}^\dag_j \hat{c}_{j\beta}\hat{P}_j\ket{0}\\
\end{split}
\end{equation}
where the last equation employs the Gutzwiller approximation. The site index $i$ will be dropped for simplicity once in need hereafter.

Now, 
\begin{equation}
\begin{split}
\hat{P}^\dag_i \hat{c}^\dag_{i\alpha}\hat{P}_i &=\sum_{I_1,I_2,I_3,I_4}  \lambda^\dag_{I_2I_1} S^\dag_{\alpha,I_3I_1} \lambda_{I_3I_4} \ket{I_2}\bra{I_4}   \\
\end{split}
\end{equation}
where $S_{\alpha,I_1I_2} = \bra{I_1}\hat{c}_{\alpha}\ket{I_2}$ and 
\begin{equation}
\begin{split}
\ket{I_2}\bra{I_4}  & =  \sum_\gamma \ket{I_2}\bra{I_2}\hat{c}^\dag_{\gamma}\ket{I_4}\bra{I_4}\\
&= \sum_{\gamma} \sqrt{\hat{l}_{\alpha}}S_{\gamma,I_4I_2}\sqrt{\hat{l}_{\alpha}}\hat{c}^\dag_{\gamma}  \\
&= \sum_{\gamma} \sqrt{\frac{\ket{I_2}\bra{I_2}}{\hat{n}_\alpha }}S_{\gamma,I_4I_2}\sqrt{\frac{\ket{I_4}\bra{I_4}}{1-\hat{n}_\alpha}}\hat{c}^\dag_{\gamma} 
\end{split}
\end{equation}
with 
\begin{equation}
\hat{l}_{\gamma} = \frac{\ket{I_2}\bra{I_2}}{\hat{n}_\gamma} = \frac{\ket{I_4}\bra{I_4}}{1-\hat{n}_\gamma}
\end{equation}
When evaluated by the non-interacting ground state $\ket{0}$, $\ket{I}\bra{I}$ will be replaced by $m^0_{I}$ and $\hat{n}_\alpha$ by $n^0_\alpha$.

Thus, 
\begin{equation}
\begin{split}
\hat{P}^\dag_i \hat{c}^\dag_{i\alpha}\hat{P}_i &=  \sum_{\alpha,I_1,I_2,I_3,I_4}\frac{\sqrt{m^0_{I_2}}\lambda^\dag_{I_2I_1}S^\dag_{\alpha,I_1I_3}\lambda_{I_3I_4}\sqrt{m^0_{I_4}}S_{\gamma,I_4I_2}}{\sqrt{n^0_\gamma(1-n^0_{\gamma})}} \hat{c}^\dag_{\gamma} \\
&= \sum_\gamma \frac{Tr \boldsymbol{\phi}^\dag \boldsymbol{S}^\dag_{\alpha}\boldsymbol{\phi} \boldsymbol{S}_\gamma}{\sqrt{n^0_\gamma(1-n^0_\gamma)}} \hat{c}^\dag_{\gamma} \\
&\equiv \sum_\gamma \mathcal{R}^\dag_{\alpha\gamma} \hat{c}^\dag_{\gamma}\\
\end{split}
\end{equation}

Similarly, 
\begin{equation}
\hat{P}^\dag_j \hat{c}_{j\beta}\hat{P}_j =  \sum_\gamma \frac{Tr \boldsymbol{\phi}^\dag \boldsymbol{S}_{\beta}\boldsymbol{\phi} \boldsymbol{S}^\dag_\delta}{\sqrt{n^0_\delta(1-n^0_\delta)}}\hat{c}_{\gamma}  \equiv \sum_\delta \mathcal{K}_{\delta\beta}\hat{c}_{\delta}
\end{equation}

And it's easy to prove that $\boldsymbol{\mathcal{R}} = \boldsymbol{\mathcal{K}}$:

\begin{equation}
\begin{split}
(\mathcal{R}^\dag_{\beta\delta})^* &= \frac{Tr(\boldsymbol{\phi}^\dag \boldsymbol{S}^\dag_\beta \boldsymbol{\phi} \boldsymbol{S}_\delta)^*}{\sqrt{n^0_\delta(1-n^0_\delta)}} =  \frac{Tr(\boldsymbol{\phi}^\dag \boldsymbol{S}^\dag_\beta \boldsymbol{\phi} \boldsymbol{S}_\delta)^\dag}{\sqrt{n^0_\delta(1-n^0_\delta)}}=\frac{Tr(\boldsymbol{\phi}^\dag \boldsymbol{S}^\dag_\beta \boldsymbol{\phi} \boldsymbol{S}^\dag_\delta)^\dag}{\sqrt{n^0_\delta(1-n^0_\delta)}} \\
&=\mathcal{K}_{\delta\beta}
\end{split}
\end{equation}

\section{PPDs of Fermi part}
Several key PPDs which are essentials for the PDs in the text will be present here. Some other second order PPDs are easily derived through the chain rule and will not be shown here.
\begin{enumerate}
\item The Fermi-Dirac distribution and its derivative
\begin{equation}
f_{n\boldsymbol{k}} = 
\left\{ 
\begin{split}
0  & \qquad  \text{if $e^{\frac{E_{n\boldsymbol{k}}}{k_BT}} > 20$} \\
 \frac{1}{e^{\frac{E_{n\boldsymbol{k}}}{k_BT}}+1}   &\qquad \text{if $e^{\frac{E_{n\boldsymbol{k}}}{k_BT}} > 20$} \\
1  &\qquad  \text{if $e^{\frac{E_{n\boldsymbol{k}}}{k_BT}} <- 20$} \\
\end{split}
\right.
\end{equation}
\begin{equation}
\frac{\partial f_{n\boldsymbol{k}}}{\partial E_{n\boldsymbol{k}} } = 
\left\{ 
\begin{split}
0  & \qquad  \text{if $e^{\frac{E_{n\boldsymbol{k}}}{k_BT}} > 20$} \\
- \frac{e^{\frac{E_{n\boldsymbol{k}}}{k_BT}}}{k_BT(e^{\frac{E_{n\boldsymbol{k}}}{k_BT}}+1)^2}   &\qquad \text{if $e^{\frac{E_{n\boldsymbol{k}}}{k_BT}} > 20$} \\
0  &\qquad  \text{if $e^{\frac{E_{n\boldsymbol{k}}}{k_BT}} <- 20$} \\
\end{split}
\right.
\end{equation} 

\item $\frac{\partial_0 E^{kin}}{\partial_0 \boldsymbol{X}} (\boldsymbol{X}= \boldsymbol{\lambda^F}, \boldsymbol{\mathcal{R}})$

\begin{equation}
\begin{split}
\frac{\partial_0 E^{kin}}{\partial_0 \boldsymbol{X}} &= \frac{\partial_0 \sum_{n\boldsymbol{k}}w(\boldsymbol{k}) f_{n\boldsymbol{k}}E_{nk}}{\partial_0 \boldsymbol{X}}\\
&= \sum_{n\boldsymbol{k}} w(\boldsymbol{k}) \frac{d f_{n\boldsymbol{k}}}{d E_{n\boldsymbol{k}}}\frac{\partial_0 E_{n\boldsymbol{k}}}{\partial_0 \boldsymbol{X}}E_{n\boldsymbol{k}} +  \sum_{n\boldsymbol{k}}w(\boldsymbol{k}) f_{n\boldsymbol{k}}\frac{\partial_0 E_{n\boldsymbol{k}}}{\partial_0 \boldsymbol{X}}  
\end{split}
\end{equation}
where $w(\boldsymbol{k})$ is the weight of $\boldsymbol{k}$ point.
\item $\frac{\partial_0 n^F}{\partial_0 \boldsymbol{X}} (\boldsymbol{X}= \boldsymbol{\lambda^F}, \boldsymbol{\mathcal{R}})$

\begin{equation}
\begin{split}
n^F_\alpha &= \sum_{n\boldsymbol{k}}w(\boldsymbol{k})f_{n\boldsymbol{k}}\big|(\sum_{m}S^{\boldsymbol{k},*}_{\alpha,m}\psi^{\boldsymbol{k},*}_{mn} )\big|^2 \\
\frac{\partial_0 n^F_\alpha}{\partial_0 \boldsymbol{X}} &= \sum_{n\boldsymbol{k}}w(\boldsymbol{k})\big(\frac{d f_{n\boldsymbol{k}}}{d E_{n\boldsymbol{k}}} \frac{\partial_0 E_{n\boldsymbol{k}}}{\partial_0 \boldsymbol{X}}\big|(\sum_{m}S^{\boldsymbol{k},*}_{\alpha,m}\psi^{\boldsymbol{k},*}_{mn} ) \big|^2 \\
&+ f_{n\boldsymbol{k}}\big|(\sum_{m}S^{\boldsymbol{k}}_{\alpha,m}\frac{\partial_0 \psi^{\boldsymbol{k}}_{mn}}{\partial_0 \boldsymbol{X}})\big|^2 \big)
\end{split}
\end{equation}

\item $\frac{\partial_0 \boldsymbol{\chi}}{\partial_0 \boldsymbol{X}} (\boldsymbol{X}= \boldsymbol{\lambda^F}, \boldsymbol{\mathcal{R}})$
\begin{equation}
\begin{split}
\frac{\partial_0 \boldsymbol{\chi}}{\partial_0 \boldsymbol{\lambda^F}}  &= \frac{\partial_0 }{\partial_0 \boldsymbol{X}}\frac{\partial_0 E^F}{\partial_0 \boldsymbol{\mathcal{R}}} - (\frac{\partial_0 }{\partial_0 \boldsymbol{X}} \frac{\partial_0 E^F}{\partial_0 \boldsymbol{\lambda^F}})(\frac{\partial_0 \boldsymbol{n^F}}{\partial_0 \boldsymbol{\lambda^F}})^{-1}\frac{\partial_0 \boldsymbol{X}}{\partial_0 \boldsymbol{\mathcal{R}}}\\
& + \frac{\partial_0 E^F}{\partial_0 \boldsymbol{\lambda^F}}(\frac{\partial_0 \boldsymbol{n^F}}{\partial_0 \boldsymbol{\lambda^F}})^{-1}(\frac{\partial_0 }{\partial_0 \boldsymbol{\lambda^F}}\frac{\partial_0 \boldsymbol{n^F}}{\partial_0 \boldsymbol{\lambda^F}} )(\frac{\partial_0 \boldsymbol{n^F}}{\partial_0 \boldsymbol{\lambda^F}})^{-1}\frac{\partial_0 \boldsymbol{n^F}}{\partial_0 \boldsymbol{\mathcal{R}}} \\
&-\frac{\partial_0 E^F}{\partial_0 \boldsymbol{\lambda^F}}(\frac{\partial_0 \boldsymbol{n^F}}{\partial_0 \boldsymbol{\lambda^F}})^{-1}\frac{\partial_0 }{\partial_0  \boldsymbol{X}}\frac{\partial_0 \boldsymbol{n^F}}{\partial_0 \boldsymbol{\mathcal{R}}}
\end{split}
\end{equation}
where we have used the equation $\frac{\partial \boldsymbol{A}^{-1}}{\partial a} = - \boldsymbol{A}^{-1}\frac{\partial \boldsymbol{A}}{\partial a}\boldsymbol{A}^{-1}$ ($\boldsymbol{A}$ is a matrix and $a$ is its variable).

\item $\frac{\partial_0 \ket{n\boldsymbol{k}}}{\partial_0 \boldsymbol{X}} (\boldsymbol{X}=\boldsymbol{\lambda^F},\boldsymbol{\mathcal{R}})$ \\
The derivative of wave functions with respect to any variable is always derived through the first perturbation theory.
\begin{equation}
\frac{\partial_0 \ket{n\boldsymbol{k}}}{\partial_0 \boldsymbol{X}} = \sum_{m\ne n} \frac{\bra{m\boldsymbol{k}}\frac{\delta \boldsymbol{H}_{\boldsymbol{k}}^F}{\delta \boldsymbol{X}}\ket{n\boldsymbol{k}}}{E_{n\boldsymbol{k}} - E_{m\boldsymbol{k}}} \ket{m\boldsymbol{k}}
\end{equation}
\end{enumerate}
From the formula above, it's obviously shown that the computation complexity is $O(N_{band}^5)$ which could be reduced to $O(N_{band}^4)$ if the part $\bra{m\boldsymbol{k}}\frac{\delta \boldsymbol{H}_{\boldsymbol{k}}^F}{\delta \boldsymbol{X}}\ket{n\boldsymbol{k}}$ is calculated and stored ahead of the calculation. It is a classic case of saving time by storage space (memory).

\section{PPDs of Bose part}
There are some techniques for the linear response theories here due to the general eigenvalue problem of the Bose part,
\begin{equation}
\boldsymbol{H}^B \boldsymbol{a}^T = E^G \mathbb{F} \boldsymbol{a}^T 
\end{equation}
In order to easily utilize the existing perturbation formula, the general eigenvalue problem is modified as 
\begin{equation}
\tilde{\boldsymbol{H}}^B \tilde{\boldsymbol{a}}^T = E^G  \tilde{\boldsymbol{a}}^T
\end{equation}
with 
\begin{equation}
\begin{split}
\tilde{\boldsymbol{H}}^B &= \mathbb{F}_{1 \over 2}^{-1} \boldsymbol{H}^B  \mathbb{F}_{1 \over 2}^{-1} \\
\tilde{\boldsymbol{a}}^T &= \mathbb{F}_{1 \over 2} \boldsymbol{a}^T \\
\mathbb{F}_{1 \over 2} &= \boldsymbol{U} \mathbb{F}_{{1 \over 2},diag} \boldsymbol{U}^{-1} \\
\end{split}
\end{equation}
where $U$ is the unitary matrix to diagonalize the $\mathbb{F}$ matrix. And $\mathbb{F}_{{1 \over 2},diag}$ is the square root of the diagonal eigenvalue matrix of $\mathbb{F}$.

\begin{enumerate}
\item $\frac{\partial_0 \boldsymbol{n^B}}{\partial_0 \boldsymbol{X}} (\boldsymbol{X}=\boldsymbol{\lambda^B},\boldsymbol{\chi})$ \\
\begin{equation}
\begin{split}
\frac{\partial_0 n^B_\alpha}{\partial_0 \boldsymbol{X}}  &= \frac{\partial_0 \boldsymbol{a}}{\partial_0 \boldsymbol{X}}\mathbb{N}_{\alpha} \boldsymbol{a}^T + \boldsymbol{a} \mathbb{N}_{\alpha} \frac{\partial_0 \boldsymbol{a}^T}{\partial_0 \boldsymbol{X}} \\
&= 2 \boldsymbol{a} \mathbb{N}_{\alpha}\frac{\partial_0 \boldsymbol{a}^T}{\partial_0 \boldsymbol{X}}
\end{split}
\end{equation}

\item $\frac{\partial_0 \boldsymbol{n^B}}{\partial_0 \boldsymbol{n^0}}$ \\
\begin{equation}
\begin{split}
\frac{\partial_0 n^B_\alpha}{\partial_0 \boldsymbol{n^0}}  &= \frac{\partial_0 \boldsymbol{a}}{\partial_0 \boldsymbol{n^0}}\mathbb{N}_{\alpha} \boldsymbol{a}^T + \boldsymbol{a} \mathbb{N}_{\alpha} \frac{\partial_0 \boldsymbol{a}^T}{\partial_0 \boldsymbol{n^0}} + \boldsymbol{a} \frac{d \mathbb{N}_\alpha}{d \boldsymbol{n^0}} \boldsymbol{a}^T \\
&= 2 \boldsymbol{a} \mathbb{N}^{\alpha}\frac{\partial_0 \boldsymbol{a}^T}{\partial_0 \boldsymbol{n^0}} + \boldsymbol{a} \frac{d \mathbb{N}_\alpha}{d \boldsymbol{n^0}} \boldsymbol{a}^T 
\end{split}
\end{equation}

\item $\frac{\partial_0 \boldsymbol{\mathcal{R}}_{n+1}}{\partial_0 \boldsymbol{X}} (\boldsymbol{X}=\boldsymbol{\lambda^B},\boldsymbol{\chi})$ \\ 
Similiarly, 
\begin{equation} 
\frac{\partial_0 \boldsymbol{\mathcal{R}}_{n+1}}{\partial_0 \boldsymbol{X}}  = 2 \boldsymbol{a} \mathbb{R}\frac{\partial_0 \boldsymbol{a}^T}{\partial_0 \boldsymbol{X}}
\end{equation}

\item $\frac{\partial_0 \boldsymbol{\mathcal{R}}_{n+1}}{\partial_0 \boldsymbol{n^0}}$ \\ 
Similiarly, 
\begin{equation} 
\frac{\partial_0 \boldsymbol{\mathcal{R}}_{n+1}}{\partial_0 \boldsymbol{n^0}}  = 2 \boldsymbol{a} \mathbb{R}\frac{\partial_0 \boldsymbol{a}^T}{\partial_0 \boldsymbol{n^0}} + \boldsymbol{a} \frac{d \mathbb{R}}{d \boldsymbol{n^0}}\boldsymbol{a}^T
\end{equation}

\item $\frac{\partial_0 E^a}{\partial_0 \boldsymbol{X}} (\boldsymbol{X}=\boldsymbol{\lambda^B},\boldsymbol{\chi})$ \\ 
Similiarly, 
\begin{equation} 
\frac{\partial_0 E^a}{\partial_0 \boldsymbol{X}}  = 2 \boldsymbol{a} \mathbb{L}\frac{\partial_0 \boldsymbol{a}^T}{\partial_0 \boldsymbol{X}}
\end{equation}

\item $\frac{\partial_0 E^a}{\partial_0 \boldsymbol{n^0}}$ \\ 
Similiarly, 
\begin{equation} 
\frac{\partial_0 E^a}{\partial_0 \boldsymbol{n^0}}  = 2 \boldsymbol{a} \mathbb{L}\frac{\partial_0 \boldsymbol{a}^T}{\partial_0 \boldsymbol{n^0}} + \boldsymbol{a} \frac{d \mathbb{L}}{d \boldsymbol{n^0}}\boldsymbol{a}^T
\end{equation}

\item $\frac{\partial_0 \boldsymbol{a}^T}{\partial_0 \boldsymbol{X}} (\boldsymbol{X}=\boldsymbol{\lambda^B},\boldsymbol{\chi})$\\
\begin{equation}
\frac{\partial_0 \boldsymbol{a}^T}{\partial_0 \boldsymbol{X}}  = \mathbb{F}_{1 \over 2}^{-1}\frac{\partial_0  \tilde{\boldsymbol{a}}^T}{\partial_0 \boldsymbol{X}}
\end{equation}

\item $\frac{\partial_0 \boldsymbol{a}^T}{\partial_0 \boldsymbol{n^0}} $\\
\begin{equation}
\frac{\partial_0 \boldsymbol{a}^T}{\partial_0 \boldsymbol{n^0}}  = \mathbb{F}_{1 \over 2}^{-1}\frac{\partial_0  \tilde{\boldsymbol{a}}^T}{\partial_0 \boldsymbol{n^0}}  - \mathbb{F}^{-1}_{1 \over 2} \frac{d \mathbb{F}_{1 \over 2}}{d \boldsymbol{n^0}} \boldsymbol{a}^T
\end{equation}

\item $\frac{\partial_0 a_m}{\partial_0 \boldsymbol{X}} $ \\
\begin{equation}
 \frac{\partial_0 \tilde{a}_m}{\partial_0 \boldsymbol{X}} = \sum_{l \ne 1}  \frac{\tilde{\boldsymbol{a}}^{l} \frac{\partial_0 \tilde{\boldsymbol{H}}^a}{\partial_0 \boldsymbol{X}} \tilde{\boldsymbol{a}}^T}{E^G-E^G_{l}} \tilde{a}_m^l
\end{equation}

where $\tilde{\boldsymbol{a}}^{l}$ and $E^G_{l}$ is the wave function and energy of the excited states of Bose eigenvalue problem.

\item $\frac{\partial_0 \tilde{\boldsymbol{H}}^B}{\partial_0 \lambda^B} $ \\
\begin{equation}
 \frac{\partial_0 \tilde{H}^B}{\partial_0 \lambda^B_\alpha}  = - \mathbb{F}^{-1}_{1 \over 2} \mathbb{N}_\alpha \mathbb{F}^{-1}_{1 \over 2}
\end{equation}

\item $\frac{\partial_0 \tilde{\boldsymbol{H}}^B}{\partial_0 \boldsymbol{\chi}} $ \\

\begin{equation}
\frac{\partial_0 \tilde{\boldsymbol{H}}^B}{\partial_0 \boldsymbol{\chi}} = \mathbb{F}^{-1}_{1 \over 2} \mathbb{R} \mathbb{F}^{-1}_{1 \over 2}
\end{equation}

\item $\frac{\partial_0 \tilde{\boldsymbol{H}}^B}{\partial_0 \boldsymbol{n^0}}$ \\
\begin{equation}
\begin{split}
\frac{\partial_0 \tilde{\boldsymbol{H}}^B}{\partial_0 \boldsymbol{n^0}} &= 2 \mathbb{F}^{-1}_{1 \over 2} \boldsymbol{H}^B \frac{d  \mathbb{F}^{-1}_{1 \over 2}}{d \boldsymbol{n^0}} + \mathbb{F}^{-1}_{1 \over 2} \frac{d \boldsymbol{H}^B}{d \boldsymbol{n^0}} \mathbb{F}^{-1}_{1 \over 2}
\end{split} 
\end{equation}

\item $\frac{d \mathbb{R}}{d \boldsymbol{n^0}}$ \\
From the PPDs above, the most basic one of the derivatives of Hamiltonian are trivial except for that with respect to the variational density matrix $\boldsymbol{n^0}$. The key part of that is the derivative of $\mathbb{R}$ with respect to $\boldsymbol{n^0}$.
\begin{equation}
\begin{split}
\frac{d \mathbb{R}_{l_1l_2}^{\delta\beta}}{d n^0_\alpha} =&  \frac{d M^{S,\delta\beta}}{d n^0_\alpha} [n^0_\delta(1-n^0_\delta)]^{-{1 \over 2}} \\
&- {1 \over 2}M^{S,\delta\beta}  [n^0_\delta(1-n^0_\delta)]^{-{3 \over 2}}(1-2n^0_\delta) \delta_{\delta\alpha}
\end{split}
\end{equation}
where the superfix of $M^S$ means the symmetrization process of $\mathbb{R}$ with the definition 
\begin{equation}
\begin{split}
M^{S,\delta\beta}_{mn} &= {1 \over 2} (M^{\delta\beta}_{mn} + M^{\delta\beta}_{nm}) \\
M^{\delta\beta}_{mn} &= Tr\boldsymbol{\phi}_m^\dag \boldsymbol{S}_\beta \boldsymbol{\phi}_n \boldsymbol{S}^\dag_\delta
\end{split}
\end{equation}
Thus the vital part turns to the derivative of  $M^{\delta\beta}_{mn}$ with respect to $\boldsymbol{n^0}$
\begin{equation}
\begin{split}
\frac{d M^{\delta\beta}_{mn}}{d n^0_\alpha} =& \sum_{IJ}{1 \over 2}\big( \frac{1}{\sqrt{m^0_I}} \frac{d m^0_I}{d n^0_\alpha} (PSP)^{m\beta n}_{IJ}\sqrt{m^0_J}S^\dag_{\delta,JI} \\
&+ \sqrt{m^0_I} (PSP)^{m\beta n}_{IJ} \frac{1}{\sqrt{m^0_J}} \frac{\delta m^0_J}{\delta n^0_\alpha}S^\dag_{\delta,JI} \big)
\end{split}
\end{equation}
with $(PSP)^{m\beta n}_{IJ} = \sum_{I'J'} \phi^{m \dag}_{II'} S_{\beta,I'J'} \phi^{n}_{J'J}$. This part is the most time consuming part which need to be accelerated.

\end{enumerate}

\bibliographystyle{elsarticle-num}
\bibliography{ref}

\end{document}